\documentclass[twocolumn]{autart}    %

\usepackage{graphicx}          %
\usepackage{epstopdf}

\usepackage{url}            %
\usepackage{booktabs}       %
\usepackage{amsfonts,amsmath,amssymb,mathrsfs}       %
\usepackage{bbm}
\usepackage{cite}
\usepackage{mathtools,xparse}
\usepackage[makeroom]{cancel}
\usepackage[acronym]{glossaries}
\usepackage{color}
\usepackage{balance}

\usepackage{algorithm}
\usepackage{algpseudocode}
\usepackage{subcaption}
\usepackage{nicefrac}       %
\usepackage{xcolor}         %
\usepackage{graphicx}
\usepackage[prependcaption,colorinlistoftodos]{todonotes}
\usepackage{multirow}
\usepackage{booktabs,tabularx}
\usepackage{cases}
\usepackage{hyperref}
\usepackage{varwidth}
\usepackage{tikz}
\usetikzlibrary{fit,arrows,calc,positioning,shapes.geometric}

\usepackage{cuted} %
\newcommand*\circled[1]{\tikz[baseline=(char.base)]{
        \node[shape=circle,draw,minimum size=4mm, inner sep=0pt] (char)
        {\rule[-3pt]{0pt}{\dimexpr2ex+2pt}#1};}}

\newcommand{\R}{\mathbb{R}}

\newcommand{\N}{\mathbb{N}}

\newcommand{\mc}[1]{\mathcal{#1}}

\newcommand{\real}{\mathbb{R}}
\newcommand{\eg}{\emph{e.g., }}

\newcommand{\eqdef}{\coloneqq}

\newcommand{\conv}{\mathrm{conv}}

\newcommand{\rank}{\mathrm{rank}}

\newcommand{\sat}{\mathrm{sat}}
\newcommand{\sign}{\mathrm{sign}}
\newcommand{\trace}{\mathrm{trace}}

\newcommand{\bsone}{\boldsymbol{1}}

\newtheorem{theorem}{Theorem}[section]

\newtheorem{lemma}[theorem]{Lemma}

\newtheorem{remark}[theorem]{Remark}

\newcommand{\qedsymbol}{\hfill$\blacksquare$}
\usepackage{dsfont} 

\makeglossaries
\newacronym{CLF}{CLF}{control Lyapunov function}
\newacronym{CEGIS}{CEGIS}{counterexample guided inductive synthesis}
\newacronym{LMI}{LMI}{linear matrix
	inequality}
\newacronym{LDI}{LDI}{linear difference inclusion}
\newacronym{SMT}{SMT}{satisfiability modulo theory}
\newacronym{AFTC}{AFTC}{active fault-tolerant control}
\newacronym{PFTC}{PFTC}{passive fault-tolerant control}
\newacronym{AUV}{AUV}{autonomous underwater vehicle}
\newacronym{FDI}{FDI}{fault detection and isolation}
\newacronym{ANN}{ANN}{artificial neural network}
\newacronym{SDP}{SDP}{semidefinite program}
\newacronym{IS-sat}{IS-sat}{inductive synthesis saturated control}
\newacronym{wrt}{w.r.t.\@}{with respect to}
\newacronym{RHS}{RHS}{right-hand side}

\makeglossaries

\begin{document}

\begin{frontmatter}

\title{Fault-tolerant control of nonlinear systems: An inductive synthesis approach} %

\thanks[footnote_corr]{Corresponding author.}

\author[Daniele Masti]{Daniele Masti}\ead{daniele.masti@gssi.it},    %
\author[Davide Grande]{Davide Grande\thanksref{footnote_corr}}\ead{ucemdg0@ucl.ac.uk, grande.rdev@gmail.com},  %
\author[Andrea Peruffo]{Andrea Peruffo},  %
\author[Filippo Fabiani]{Filippo Fabiani}\ead{filippo.fabiani@imtlucca.it}  %

\address[Daniele Masti]{Gran Sasso Science Institute, Viale F. Crispi 7, 67100, L'Aquila, IT}  %
\address[Davide Grande]{University College London, Gower St, WC1E 6BT London, UK }             %

\address[Andrea Peruffo]{Delft University of Technology, Mekelweg 5, 2628 CD Delft, NL}        %

\address[Filippo Fabiani]{IMT School for Advanced Studies Lucca, Piazza San Francesco 19, 55100 Lucca, IT}        %


\begin{abstract}                          
Actuator faults heavily affect the performance and stability of control systems, an issue that is even more critical for systems required to operate autonomously under adverse environmental conditions, such as unmanned vehicles. To this end, \gls{PFTC} systems can be employed, namely fixed-gain control laws that guarantee stability both in the nominal case and in the event of faults. In this paper, we propose a \gls{CEGIS}-based approach to design reliable \gls{PFTC} policies for nonlinear control systems affected by partial, or total, actuator faults. Our approach enjoys finite-time convergence guarantees and extends available techniques by considering nonlinear dynamics with possible fault conditions. Extensive numerical simulations illustrate how the proposed method can be applied to realistic operational scenarios involving the velocity and heading control of \glspl{AUV}. Our \gls{PFTC} technique exhibits comparatively low synthesis time (i.e. minutes) and minimal computational requirements, which render it is suitable for embedded applications with limited availability of energy and onboard power resources. 
\end{abstract}

\end{frontmatter}

\glsresetall
\section{Introduction}

Unmanned vehicles are usually employed over long periods of time in unstructured or adverse environments to accomplish several tasks, \eg data collection, environmental monitoring and patrolling. 
During the mission, these robots may be subject to actuators or sensors faults. In words, a fault amounts to an undesired change in the dynamics of a signal of a sensor or an actuator, which does not compromise the entire functionality of the overall system \cite{ducard2009fault}. A fault can be worked around so that the system can still accomplish its original task, even with a certain degree of performance degradation \cite{blanke2006diagnosis}. 
Owing to the persistent exposure of the actuators to the surrounding environment, 
actuator faults represent the most common issue in operations involving modern underwater robots \cite{liu2023review}. 

Dealing with actuator faults is traditionally addressed via two nearly complementary approaches. On the one hand, \gls{FDI} algorithms \cite{caiti2015enhancing,fabiani2016hybrid} are designed to detect the presence of non-nominal operating conditions, isolating the faulty component and estimating the severity of the fault occurrence. As a consequence, the system's control policy is either rescheduled, or a completely new control architecture, tailored for the specific faulty behaviour, is employed. These approaches fall within the \gls{AFTC} techniques. On the other hand, \gls{PFTC} methods consider a set of possible faulty dynamics and the design of a fixed-gain control law, guaranteeing to preserve closed-loop stability over the whole range of faulty behaviours.
\gls{PFTC} is frequently compared to solving a robust stabilisation problem, thereby requiring the simultaneous stabilisation of a plethora of dynamics catering for the nominal plant, and for the set of operating modes under faults.

\textbf{Related work: }
Leveraging the tight interaction between an \gls{FDI} block with a control law scheduler or gain adaptive engine, \gls{AFTC} methods offer the possibility to optimise a system's closed-loop performance in diverse operating conditions. On the other hand, \gls{AFTC} techniques are typically computationally expensive, require accurate model descriptions, and need time to estimate the fault location and its severity, introducing a critical delay that might lead to instability \cite{Verhaegen2010}. \gls{PFTC} architectures, instead, eliminate the need for monitoring sensors and are computationally inexpensive to implement, as they rely on fixed-gain controllers. When either extensive actuator sensoring, control design or online deployment possibilities are limited due to power, costs, or complexity constraints, \gls{PFTC} represents the most effective control option.
On the other hand, \glspl{PFTC} result in more conservative control performance, even under nominal operating conditions. However, in safety-critical applications, or when the human-in-the-loop intervention is unfeasible, \eg \glspl{AUV} operating under the polar ice caps \cite{webster2015towards}, reliability and safety need to be prioritised \cite{kaminer1991control}.
Available \gls{PFTC} methods exploit both linear or nonlinear control theory. Linear methods usually stem from $\mathcal{H}_2$ and $\mathcal{H}_\infty$-control synthesis~\cite{blanke2006diagnosis}; 
analytical nonlinear techniques instead rely on Lyapunov theory, for instance by designing a control law for the fault-free model, and adding extra terms catering for changes and/or faults in the actuator dynamics \cite{benosman2009passive,spooner2004stable}.

Computer-aided design methods have been recently employed in the design of control systems. One of the most prominent examples refers to the \gls{CEGIS} technique, which consists of a loop between two components. Specifically, a \emph{learner} is in charge of designing a candidate control solution, which is then passed to a \emph{verifier} that checks whether the candidate solution is valid over the whole state domain. 
\gls{CEGIS}-based schemes have recently demonstrated their potential in the design of \glspl{CLF}, or to extend the capabilities of traditional robust control approaches \cite{solar2006combinatorial,grande2023augmented,berger2022learning,chang2019neural,dai2020counter,Abate2023Neural,masti2023counter}. With this regard, to overcome the intrinsic computational complexity resulting in the verifier's task, recent works have focused on \gls{SMT}-solvers as verification engines. 
For instance, ANLC \cite{grande2023augmented} and Fossil \cite{edwards2023fossil} employ \gls{SMT} to verify the closed-loop stability of continuous-time systems, thus offering an alternative approach to standard techniques based on mixed-integer verification \cite{dai2020counter,wu2024neural} or on Lipschitz-based optimization~\cite{he2022approximate,masti2023counter}. Despite the attractiveness of the automatic synthesis of correct-by-design \glspl{CLF}, \gls{CEGIS}-based methods are not usually guaranteed to converge.
Besides introducing heuristics to increase the successful synthesis rate \cite{grande2023augmented}, little work has been done to theoretically guarantee algorithmic termination.
As we were finalising this manuscript, we became aware of a recent contribution \cite{hsieh2025certifying} that presents an approach related to ours. However, the work in \cite{hsieh2025certifying} involves a significantly different technical analysis. In fact, it leverages Lipschitz continuity to learn Lyapunov functions for \emph{unknown} dynamical systems, while relying on \gls{SMT} for verification purposes.

\textbf{Contribution: }
We  extend the \gls{CEGIS}-based approach in \cite{masti2023counter} in four different directions: 
$i)$ we consider \emph{nonlinear} control systems; 
$ii)$ the control design is tailored to \emph{fault-tolerant} policies; 
$iii)$ it is supported by theoretical results ensuring both the validity of our approach, and the finite-step convergence of the algorithm; 
$iv)$ control inputs can account for actuator \emph{saturations}, extending the applicability to real-world physical systems. 
In contrast to \cite{grande2023systematic,grande2024passive}, our method enjoys a specialised verification engine, which guarantees finite-time convergence and improves scalability, making this technique applicable to complex cyber-physical systems.  
We define our method as \gls{IS-sat}.

The rest of the paper is organised as follows: in \S \ref{sec:problem_definition} we formally introduce the considered \gls{PFTC} problem; in \S \ref{sec:uncertain-cegis} we detail our method while in \S \ref{sec:experiments} we show its experimental effectiveness over an \gls{AUV} benchmark.

\subsubsection*{Notation}
$\mathbf{1}_n$ is an $n$-dimensional vector of 1. 
$\mathds{1}(\varphi)$ denotes the indicator of function of $\varphi$, i.e., $\mathds{1}(\varphi)=1$ when the condition $\varphi$ is true, $0$ otherwise. The operator $\mathrm{sat}_{\mathscr U}(u)$ denotes the componentwise saturation function, i.e., for $u\in\R^p$, 
\begin{align}
\mathrm{sat}_{\mathscr U}(u) = [ \ldots, \  \sign(u_i) \cdot \min\{ u_{M i}, |u_i| \} , \ \ldots ],
\label{eq:satDef}
\end{align}
with sign function $\sign(u_i)$ and $i$-th saturation threshold $u_{Mi}$. The operator $\conv\{\cdot\}$ denotes the convex hull of its arguments.

\section{Problem formulation and preliminaries}
\label{sec:problem_definition}

\subsection{System description}

Let us consider the following control-affine nonlinear model: 
\begin{equation}
	\label{eq:system-model}
	{x}(t+1) = f(x(t)) + g(x(t), \phi(t)) u(t),
\end{equation}
where $t \in \N_+$ denotes the discrete-time index, $x(t) \in \mathscr{D} \subseteq \R^n$ is the vector of state variables, $u(t) \in \mathscr{U} \subset \R^p$ is the vector of control inputs,
the mappings $f:\R^n\to\R^n$ and $g:\R^n\times [0, 1]^p \to \R^{n \times p}$ are of class $\mc C^2$ \gls{wrt} $x(t)$, $\phi(t)$,
and $\phi(t) \in [0, 1]^p$ denotes the possible operation modes of the system. While $\phi_i=1$, for all $i=1,\ldots,p$, represents the nominal operation, $\phi_i \in [0,1]$ models the loss of efficiency of actuator $i$, \eg  $\phi_i=0.6$ indicates that actuator $i$ can only work at 60\% of the nominal value. Finally, $\phi_i=0$ denotes the total fault of actuator $i$.
In this work, we assume there can be just one entry of vector $\phi$ that assumes a value in $[0,1]$. More formally we impose that $\phi(t) \in \Phi \eqdef \{\phi\in[0, 1]^p \mid \sum_{i=1}^p \mathds{1}(\phi_i=1) \geq p-1 \}$. This assumption is not restrictive and commonly adopted in, among the many example, \gls{AUV} control engineering~ \cite{caiti2015enhancing,fabiani2016hybrid,grande2024passive}, as faults are rare events. As such, multiple faults occurring during the same operation might require a complete stop of operations as the vehicles are by design not capable to cope with such unlikely scenarios. Without loss of generality we also assume the origin being an equilibrium point of~\eqref{eq:system-model}, i.e., for a given $\bar\phi\in\Phi$, there exists some control law $\bar u \in \mathscr{U}$ so that $f(0) + g(0,\bar\phi)\bar u=0$. 

For computational purposes, we finally constraint both the state variables and control inputs within polytopes $\mathscr{D}\eqdef\{ x \in \real^n \mid Lx \leq \bsone_\ell \}$ and $\mathscr{U}\eqdef\{u\in\real^p\mid|u| \leq \bar u\}$, respectively, for some $L\in\R^{\ell\times n}$, $\rank(L)=\ell$, and $\bar u\in\R^p_+$. Note that $\mathscr U$ actually coincides with a saturation on the control input, element which we take into account in the presented derivation, in line with realistic control engineering applications.

Our goal is hence to design a control law for model \eqref{eq:system-model} under all possible faults happening one at a time, i.e., under all possible values $\phi$ can take in $\Phi$, with the aim of guaranteeing the maximal region of attraction, whilst abiding the constraints imposed on both state variables and control inputs. As a byproduct of the control method we will propose, a Lyapunov function will certify the closed-loop stability.

\subsection{Reformulation as uncertain system}
\label{sec:unc_sys}

Typically, a controller for system \eqref{eq:system-model} is designed by means of robust control techniques based on, \eg  system linearisation or Lyapunov arguments. The latter offers analytical solutions for a partial actuator fault, but conventionally can not cope with the case of full loss of efficiency, namely $\phi_i=0$ \cite{benosman2009passive}. Inspired by the uncertain system literature, we reformulate the stabilisation of \eqref{eq:system-model} as the uncertain model: 
\begin{equation}
	\label{eq:model-linear-uncertain}
	{x}(t+1) =  A(t) x(t) + B(t) u(t), 
\end{equation}
where matrices $A(t) \in \R^{n\times n}$ and $B(t) \in \R^{n \times p}$ belong to:
\begin{multline}
\label{eq:def-omega}
    \Omega \eqdef 
    \left\{  
    (A,B)   \mid 
    A = \left. \frac{df}{dx}\right|_{x = \bar{x}}, 
    \right.
    \\  
    \left. 
    B = \left. \frac{dg}{dx}\right|_{\substack{x = \bar{x} \\ \phi = \bar{\phi}}}, 
    \text{ for } \bar x\in\mathscr{D}, \ \bar \phi\in\Phi
    \right\} ,
\end{multline}
which collects all the Jacobian matrices $A(\cdot)$ of the autonomous (i.e. time-invariant) dynamics in~\eqref{eq:system-model}, as well as those related to the affine control term $B(\cdot)$, which consider all possible variations due to an actuator fault through $\phi \in \Phi$. 
We will then assume that both $A(t)$ and $B(t)$ are bounded for all $x \in \mathscr{D}$ and $t \in \R_{+}$, which is a natural condition for standard nonlinear models describing physical system, \eg the motion of \glspl{AUV} %
\cite{fossen2011handbook} considered as case study in \S \ref{sec:experiments}. 
Note that, in view of the results in \cite{liu1968convergent}, guaranteeing the closed-loop stability to~\eqref{eq:model-linear-uncertain} with $(A(t), B(t)) \in \Omega$,  implies the (local, in a neighborhood of $x(t)=0$) closed-loop stability of the original system \eqref{eq:system-model}, since $\Omega$ includes all possible behaviours that \eqref{eq:system-model} may exhibit. 
In other words, designing a controller for the system in \eqref{eq:model-linear-uncertain} yields a valid fault-tolerant policy for \eqref{eq:system-model}.

To conclude, note that $\Omega$ can assume any (possibly nonconvex) shape, which in view of~\cite[Th. 4.14 and 4.22]{rudin1964principles} it is however connected and compact.
Moreover, recall that if $\Omega$ was a polytope with $s$ vertices (as in, \eg \cite{rubin2020interpolation}), it could be written as $\Omega~=~\conv\left\{(A_i, B_i)\right\}_{i=1}^s$, also guaranteeing that some $\alpha\in\R^s_+$ exists so that $A(t)=\sum_{i=1}^s \alpha_i A_i$ and $B(t)=\sum_{i=1}^s \alpha_i B_i$, for all $t\ge0$. This will be key for our subsequent developments.

\subsection{Tackling input saturation}
\label{subsec:input_saturation}

From our assumption on $\mathscr U$, we hence aim to find a \gls{PFTC} law for \eqref{eq:system-model} of the form:
\begin{align}\label{eq:sat_con}
	u(t)=\sat_{\mathscr U}(K x(t)),
\end{align}
where $K\in\R^{p\times n}$ is a linear gain to be designed, and the saturation function restricts the control actions to the set $\mathscr{U}$. 

Various solutions have been proposed to achieve such a goal. Among them, \cite{hu2001control,rubin2020interpolation} rely on \glspl{LDI} . 
First, let us introduce an ellipsoid defined as $\mathcal{E}(Q) \eqdef \{ x\in\R^n \mid x^\top Q^{-1}x \leq 1 \}$, for some $Q\succ0$ that has to be suitably designed. 
The saturation function can hence be reformulated using matrices  $\{E_j\}_{j = 1}^{2^p}$, $E_j\in\R^{p\times p}$, representing the collection of diagonal matrices with $(0,1)$-entries. Thus, the control law in \eqref{eq:sat_con} can be rewritten as: 
\begin{equation}
	\label{eq:sat-as-ldi}
	\sat_{\mathscr U}(K x) = \sum_{j=1}^{2^p} \sigma_j (E_j Kx + E^-_j Hx), 
	\ 
	\sum_{j=1}^{2^p} \sigma_j = 1, \ \sigma_j \geq 0, 
\end{equation}
where $E^-_j \eqdef I - E_j$, and $|Hx| < \bar u$ holds within the ellipsoid $\mathcal{E}(Q)$. 
This reformulation entails an extended control design problem, in which we strive to finding a (possibly maximal) invariant ellipsoid $\mathcal{E}(Q)$, along with matrices $K$, $H$. 

To this end, we introduce the three matrices $Q \succ 0$, $Y~=~KQ$, $Z = HQ$, and we impose the following \gls{LMI}, where ``$\star$'' denotes the corresponding transposed block-element to make the LHS symmetric:
\begin{subequations}
\label{eq:BemporadLMI}
\begin{multline}
	\label{eq:schur-sat-LMI-bemporad}
	\begin{bmatrix}
		\tau Q  & \star         & \star 
		\\
		0       & (1-\tau) I    & \star 
		\\
		A_i Q + B_i E_j Y + B_i E^-_j Z & 0 & Q
	\end{bmatrix}
	\succcurlyeq 0, %
	\\ 
	\text{for all }i = 1,\ldots, s, \; j = 1, \ldots , 2^p, 
\end{multline}
where $\tau \in [0, 1]$ is an hyperparameter whose meaning is discussed in detail in~\cite[\S 4.1]{rubin2020interpolation} and where the pair of matrices $(A_i,B_i)$ represent the $i$-th vertex of $\Omega$ when $\Omega$ is defined as a polytope. This consideration will be key in the definition of the learner's task. %

We then further add the \gls{LMI} conditions to satisfy state and input constraints, as follows:
\begin{equation}
	\label{eq:state-constraint-bemporad}
	\begin{bmatrix}
		1 & l_i Q
		\\
		Q l_i^\top & Q
	\end{bmatrix}
	\succcurlyeq 0, \; \text{for all } i = 1, \ldots, \ell
\end{equation}
where $l_i$ is the $i$-th row of matrix $L$, and 
\begin{equation}
	\label{eq:input-constraint-bemporad}
	\begin{bmatrix}
		\bar u_i^2 & z_i
		\\
		z_i^\top & Q
	\end{bmatrix}
	\succcurlyeq 0, \; \text{for all } i = 1, \ldots, p,
\end{equation}
\end{subequations}
where $z_i$ is the $i$-th row of matrix $Z$.
Finding a set of matrices satisfying~\eqref{eq:BemporadLMI} amounts to solving a \gls{SDP}, which is a convex optimization problem. 
In our specific \gls{PFTC} application, we will then incorporate~\eqref{eq:BemporadLMI} into an optimization problem to find the control law with the maximal invariant ellipsoid, thus guaranteeing attraction for the largest possible region contained in $\mathscr{D}$. This can be accomplished by exploiting the following lemma:
\begin{lemma}\textup{(\hspace{-.05cm}\cite{rubin2020interpolation})}
    The largest invariant ellipsoid $\mathcal{E}(Q)$ \gls{wrt} system \eqref{eq:system-model} and any saturated control $u(t)=\sat_{\mathscr U}(K x(t))$, can be computed as $K=YQ^{-1}$ by solving the \gls{SDP}: 
    \begin{equation}
	    \label{eq:bemporad-interpolation-sdp}
	        \begin{aligned}
		            &\max_{Q,Y,Z} && \quad \trace(Q)
		            \\
		            &~~\mathrm{ s.t.} && \quad \eqref{eq:schur-sat-LMI-bemporad},\eqref{eq:state-constraint-bemporad},\eqref{eq:input-constraint-bemporad}.
		        \end{aligned}
	    \end{equation}
        \hfill$\square$
\end{lemma}
Solving \eqref{eq:bemporad-interpolation-sdp} returns three matrices: $Q$ represents the invariant ellipsoid $\mathcal{E}(Q)$, $Y$ is used to find the controller $K = YQ^{-1}$, and similarly $H = ZQ^{-1}$. 

Despite its convexity, finding a solution to \eqref{eq:bemporad-interpolation-sdp} is hard due to the combinatorial number of constraints~\eqref{eq:schur-sat-LMI-bemporad}. The \gls{LDI} formulation introduces $2^p$ \glspl{LMI}, nested with $2^{v}$ constraints coming from the uncertainty set ($v$ denotes the number of uncertain parameters of the matrices $A(t)$ and $B(t)$ belonging to $\Omega$), producing a total of $2^{p + v}$ constraints. 
The \gls{SDP} in \eqref{eq:bemporad-interpolation-sdp} is hence computationally challenging even for low dimensional models. 
Moreover, to enhance generalisability of the present work, we do \emph{not} restrict the study to $\Omega$ being necessarily a polytope. Nevertheless, $\Omega$ can be over-approximated, without loss of generality, by the convex hull composed of the $2^{v}$ uncertainty constraints. %

\section{CEGIS-based synthesis of PFTC policies}
\label{sec:uncertain-cegis}

\subsection{Counterexample-guided iterative procedure}
\label{subsec:cegis_procedure}
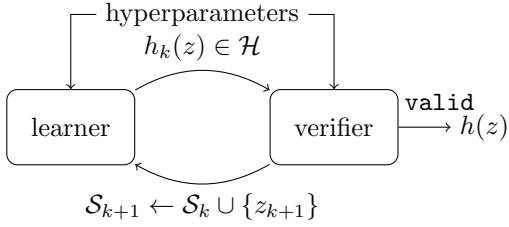
\begin{figure}[t!]
	\centering
	\usetikzlibrary{calc}

\begin{tikzpicture}[transform shape]
    \node[shape=rectangle,rounded corners, draw,minimum width=1.7cm, minimum height=1cm] (lea) at (0,0) {learner};
    \node[shape=rectangle, rounded corners, draw,minimum width=1.7cm, minimum height=1cm] (ver) at (3.5,0) {verifier};
    \draw[->, bend left] (lea) to node[above] {$h_k(z) \in \mc H$} (ver);
    \draw[->, bend left] (ver) to node[below] {$\mc S_{k+1} \leftarrow \mc S_{k} \cup \{z_{k+1}\}$} (lea);

    \node[] (prob) at (1.75,1.5) {hyperparameters};
    \draw[->] (prob.west) to (0,1.5) to (0,1) to (lea);
    \draw[->] (prob.east) to (3.5,1.5) to (ver);
    
    \node (out) at (5.5,0) {$h(z)$};
    \draw[->] (ver) to node[above,xshift=.2cm,yshift=.1cm] {\tt valid} (out);
  \end{tikzpicture}
	\caption{At every iteration $k \in \N_+$, the learner proposes a candidate function $h_k(z)$, while the verifier checks its validity through $r(h(z)) \le 0$.}
	\label{fig:cegis-loop}
\end{figure}

Recently, the \gls{CEGIS} learning paradigm has been adapted to the design of control functions for systems affected by actuator faults, \eg  \cite{grande2023systematic, grande2024passive}. As illustrated in Fig.~\ref{fig:cegis-loop}, \gls{CEGIS} relies on the adversarial interaction between a \textit{learner} and a \textit{verifier}, aiming at designing a function $h: \R^n \to \R$ from a hypothesis space $\mathcal{H}$, by exploiting a dataset
of (counter-)examples $\mc S_k$ (at the $k$-th algorithmic step). The procedure follows two iterative steps: 
\begin{enumerate}
	\item The learner takes the dataset $\mc S_k$ as input and either synthesises a function $h_k \in \mc H$ according to a given criterion, or establishes that such a function can not be synthesised;
	\item The verifier checks if  $h_k \in \Theta$, with $\Theta$ representing a verification criterion, namely $\Theta= \{h : \R^n \to \R \mid r(h(z)) \le 0, \text{ for all } z \in \mc Z\} $,
	for some $r : \R \to \R$
	and set $\mc Z \subseteq \R^n$ over which the variable $z$ takes values. The verifier can therefore take two different conclusions:
	\begin{enumerate}
		\item It finds a \textit{counterexample} $z_{k+1} \in \mc Z$ such that $r(h_k(z_{k+1}))>0$. In such a case
		$
		\mc S_{k+1} \leftarrow \mc S_k \cup \{z_{k+1}\},
		$
		and a new iteration step follows;

		\item It certifies that no counterexample exists, yielding the successful conclusion of the procedure.
		
	\end{enumerate}
\end{enumerate}

To design \gls{PFTC} policies, we propose to solve the \gls{SDP} in \eqref{eq:bemporad-interpolation-sdp} by using a set of strategically placed constraints which are added iteratively %
based on a \gls{CEGIS} approach. In the following sections, we illustrate in detail how the learner and verifier can be designed to solve a \gls{PFTC} problem.

\subsection{Learner's task}
\label{subsec:learner}

Let us consider a set $\mc S= \{(\hat A_i, \hat B_i)\}^{k}_{i=1}$  consisting of pair matrices $ (\hat A, \hat B) \in \Omega$.
At the algorithmic step $k \in \N_+$, the learner aims at solving the following reduced instance of \eqref{eq:bemporad-interpolation-sdp}:
\begin{subequations}
	\label{eq:learner-optimisation-LMI}
	\begin{align}
			&\max_{Q,Y,Z} && \quad \trace(Q)
			\\
			&~~\mathrm{ s.t. }
			&&\begin{bmatrix}
				\tau Q  & \star         & \star 
				\\
				0       & (1-\tau) I    & \star 
				\\
				A_i Q + B_i E_j Y + B_i E^-_j Z & 0 & Q
			\end{bmatrix} \succcurlyeq \varepsilon I, \notag 
   \\ 
   &&&\quad \quad \text{ for all }  i = 1, \ldots , V_k, \; j = 1, \ldots, 2^p, \label{eq:exp-Bemporad-schur}
   \\
      &&&\eqref{eq:state-constraint-bemporad},\eqref{eq:input-constraint-bemporad} \notag 
   \\
			&&&Q \preccurlyeq \eta I, \; Y \in \mc Y, \; Z \in \mc Z,	
	\end{align}
\end{subequations}
which considers only the $V_k$ vertices of the convex hull of $\mc S $.
In contrast to \eqref{eq:bemporad-interpolation-sdp}, formulation \eqref{eq:learner-optimisation-LMI} considers two additional hyperparameters $\eta \ge \varepsilon>0$ that can be chosen arbitrarily. As in~\cite{masti2023counter}, $\eta$ and $\varepsilon$ are meant to upper (respectively, lower) bound the largest (smallest) eigenvalue of matrix $Q$. 
Finally, we assume $\mc Y$ and $\mc Z$ to be nonempty, convex and compact subset of $\R^{m\times n}$.

An optimal solution to \eqref{eq:learner-optimisation-LMI} consists of a triplet $(Q^\ast, Y^\ast, Z^\ast)$, which allows us to determine, at each $k$-th iteration of the algorithm, a candidate Lyapunov function $V(x) = x^\top P_k x$ and a linear controller through the following quantities:
\begin{equation}
	\label{eq:solution-of-LMI}
	P_k = (Q^\ast)^{-1}, \; 
    K_k = Y^\ast (Q^\ast)^{-1},  \;
    H_k = Z^\ast (Q^\ast)^{-1},
\end{equation}
which are successively passed to the verification task.

\subsection{Verifier's task}
\label{subsec:verifier}
The verifier ensures that the triplet $(Q_k^\ast,Y_k^\ast,Z_k^\ast)$, provided by the learner as an optimal solution to \eqref{eq:learner-optimisation-LMI} at the $k$-th iteration of the iterative procedure, is actually valid over $\Omega$ and, hence, for all $x \in \mathscr D$ and  $\phi \in {\Phi}$ according to the discussion in \S \ref{sec:unc_sys}.
We first notice that the two conditions \eqref{eq:state-constraint-bemporad}--\eqref{eq:input-constraint-bemporad} concern solely the state and input domains and thus are satisfied by design. %
To check the validity of \eqref{eq:exp-Bemporad-schur} uniformly over $\Omega$---rather just on the considered samples---one can verify if the optimal value of the following optimization problem is positive:
\begin{equation} 
	\label{eq:verifier-optimisation}
	\lambda^* = \underset{(A, B) \in \Omega, j=1,\dots,2^p}{\min} \ \lambda_{\textrm{min}}(\Xi_k(A, B,j)),
\end{equation}
where $\lambda_{\textrm{min}}$ indicates the minimum eigenvalue of matrix $\Xi_k$, which corresponds to the block matrix in \eqref{eq:exp-Bemporad-schur}, computed using  $(Q_k^\ast, Y_k^\ast, Z_k^\ast)$, as a function of matrices $A$ and $B$ and the index $j$. In the following, We drop the dependence on the $j$ index momentarily to alleviate notation clutter.

Therefore, at iteration $k$ of the \gls{CEGIS} loop, the verifier computes $\Xi_k$ using the triplet $(Q_k^\ast, Y_k^\ast, Z_k^\ast)$  returned by the learner, and solves \eqref{eq:verifier-optimisation}. If $\lambda^* > 0$, no counterexample exists and the \gls{CEGIS} procedure terminates with output $(P_k, K_k, H_k)=( (Q^\ast)^{-1},  Y^\ast (Q^\ast)^{-1}, Z^\ast (Q^\ast)^{-1})$. Conversely, if the optimisation finds $\lambda^* \leq 0$, the verifier then adds the minimiser pair $(A^*, B^*) \eqqcolon (\hat A_{k+1},\hat B_{k+1})$ to the training set $\mc S_k$ such that the learner can compute a new candidate triplet $P_{k+1}$, $Y_{k+1}$, $Z_{k+1}$, and the process repeats. 

Solving~\eqref{eq:verifier-optimisation} is a nonconvex optimization problem that shall be solved \textit{globally} to
prove that no counterexample exists. To achieve global optimality, an idea is to exploit the fact that the minimum eigenvalue $\lambda_{\textrm{min}}(\cdot)$ of a symmetric matrix is a Lipschitz continuous map of the entries of the matrix itself:
\begin{lemma}[\hspace{-.01cm}\cite{horn1994topics}]%
	Consider two symmetric matrices $K$ and $L$ of the same dimension. Then, it holds that
	$
	| \lambda_{\mathrm{min}}(K)-\lambda_{\mathrm{min}}(K+L)| \leq || L||_{\mathrm{op}}$,
	where $|| \cdot||_{\mathrm{op}}$ is the operator norm of a matrix,  induced by the $l_2$-norm.
	\hfill$\square$
	\label{lemma:liplamdamin}
\end{lemma}
Yet, problem~\eqref{eq:verifier-optimisation} still involves optimizing over $\Omega$, which is an unknown, potentially nonconvex, set.
Even though one may know the domain of the state variables $\mathscr{D}$, inferring useful information for $\Omega$ from $\mathscr{D}$ itself is not straightforward.
If we however further assume that $A(t)$, $B(t)$ are Lipschitz continuous with respect to $x$---and recalling \eqref{eq:def-omega}---one can exploit the fact that the composition of two Lipschitz functions is itself Lipschitz, i.e., 
\begin{equation}
\begin{aligned}
\label{eq:LipSchitzGradient}
    || \left. A \right|_x - \left. A \right|_y || \leq \kappa_A ||x -y||, \text{ for all } x, y \in \mathscr{D},
    \\
    || \left. B \right|_{x, \phi} - \left. B \right|_{y, \psi} || \leq \kappa_B (||x -y|| + ||\phi - \psi || ), 
    \\
    \qquad \qquad \quad 
    \text{ for all } x, y \in \mathscr{D}, \  \phi, \psi \in \Phi, 
\end{aligned}
\end{equation}
for a matrix norm of interest and constants $\kappa_A$, $\kappa_B \geq 0$.

Since bounded linear operators preserve Lipschitz
continuity, we are then able to prove the following result:
\begin{theorem}
\label{theorem:one}
    Let $P = (Q^\ast)^{-1}$ and $K = Y^\ast (Q^\ast)^{-1}$, $H = Z^\ast (Q^\ast)^{-1},$ with $(Q^\ast, Y^\ast, Z^\ast)$ a solution of \eqref{eq:learner-optimisation-LMI}, with a given $\eta \geq \varepsilon > 0$. Then, there exists some constant $\ell = \ell(\varepsilon) \geq 0$ such that
    \begin{multline}
        \label{eq:theo-1-eigval-condition}
        \mid \lambda_{\mathrm{min}}(\Xi(A, B)) - \lambda_{\mathrm{min}}(\Xi(A + \Delta A, B + \Delta B)) \mid 
        \\ 
        \leq 
        \ell (\| \Delta A\|_{\mathrm{op}} + \|\Delta B \|_{\mathrm{op}})
    \end{multline}
    where $\Delta A$, $\Delta B$ denote any perturbation of $A$, $B$ with $(A, B) \in \Omega$. 
    \hfill$\square$
\end{theorem}
\begin{pf}
    To prove this statement we make use of similar arguments as in \cite[Th.~1]{masti2023counter}. 
    In fact, from Lemma~\ref{lemma:liplamdamin} we get
    \begin{align*}
         \mid \lambda_{\mathrm{min}}(\Xi(A, B)) -& \lambda_{\mathrm{min}}(\Xi(A + \Delta A, B + \Delta B)) \mid \\
         & \leq \begin{Vmatrix}
             0 & 0 & \Delta_{AB}\\
             0 & 0& 0\\
             \Delta_{AB}^\top &0 &0 
         \end{Vmatrix}_\mathrm{op},\end{align*}
 where $
         \Delta_{AB}\eqdef\Delta A Q + \Delta B (E_j Y + E_j^-Z), 
$
and $\| \cdot \|_\mathrm{op}$ is the matrix norm induced by the $\ell_2$ one. 
Manipulating the \gls{RHS}, we obtain: 
\begin{align*}
    \begin{Vmatrix}
             0 & 0 & \Delta_{AB}\\
             0 & 0& 0\\
             \Delta_{AB}^\top &0 &0 
         \end{Vmatrix}_\mathrm{op}&= \underset{\|x\|=1}{\mathrm{sup}} 
         \begin{Vmatrix}
        \begin{bmatrix}
             0 & 0& \Delta_{AB} \\ 0 & 0 &0 \\ \Delta_{AB}^\top &0 &0
         \end{bmatrix}
         \begin{bmatrix}
             x_1\\ x_2 \\x_3
         \end{bmatrix}
         \end{Vmatrix}_2\\
         &= \underset{\|x\|=1}{\mathrm{sup}} 
         \begin{Vmatrix}
         \begin{bmatrix}
             \Delta_{AB} x_3\\ 0 \\ \Delta_{AB}^\top x_1
         \end{bmatrix}
         \end{Vmatrix}_2\end{align*}
         \begin{align*}
         =\underset{\|x\|=1}{\mathrm{sup}}\sqrt{ \|  \Delta_{AB}\|_\mathrm{op} \|x_1\| + \| \Delta_{AB}\|_\mathrm{op} \|x_3\| }=\| \Delta_{AB}\|_\mathrm{op},
         \end{align*}
         \begin{align*}
         \| \Delta_{AB}\|_\mathrm{op} & \leq || \Delta A||_\mathrm{op}  ||Q||_\mathrm{op}+|| \Delta B||_\mathrm{op} ||E_j Y + E_j^- Z||_\mathrm{op}
         \\
         &\leq || \Delta A||_\mathrm{op} \eta +  ||\Delta B||_\mathrm{op}( || Y||_\mathrm{op} + ||Z||_\mathrm{op}).
         \end{align*}
This follows from the fact that $\|E_j\|_\mathrm{op}=1, \|E^-_j\|_\mathrm{op}=1~~\forall j$.
Moreover, since $\mathcal{Y}$ and $\mathcal{Z}$ are compact, both $Y$ and $Z$ are norm bounded, i.e. it holds that $||Y||_\mathrm{op} \leq \eta_Y$ and $||Z||_\mathrm{op} \leq \eta_Z$, for some $\eta_Y$, $\eta_Z>0$, and therefore
\begin{align*}
    \mid \lambda_{\mathrm{min}}(\Xi(A, B)) -& \lambda_{\mathrm{min}}(\Xi(A + \Delta A, B + \Delta B)) \mid &\\ & \leq || \Delta A||_\mathrm{op} \eta +  ||\Delta B||_\mathrm{op}( \eta_Y + \eta_Z).
\end{align*}
In turn, if $\eta$ is chosen such that $||Y||_\mathrm{op} \leq \frac{\eta}{2}$ and $||Z||_\mathrm{op}\leq \frac{\eta}{2}$, it holds that 
\begin{align*}
    \mid \lambda_{\mathrm{min}}(\Xi(A, B)) -& \lambda_{\mathrm{min}}(\Xi(A + \Delta A, B + \Delta B)) \mid &\\ & \leq \eta \cdot  ( || \Delta A||_\mathrm{op} + ||\Delta B||_\mathrm{op}),
\end{align*}
which concludes this proof. \qedsymbol
\end{pf}
\begin{remark}
\label{rem:JacLipSafeSet}
    By exploiting~\eqref{eq:LipSchitzGradient} we can further say that,
    if $\Delta A =A|_x-A|_y $ and $\Delta B =B|_{x, \phi} - B|_{y, \psi}$, then it holds that
    \begin{align*}
    \mid \lambda_{\mathrm{min}}(&\Xi(A, B)) - \lambda_{\mathrm{min}}(\Xi(A + \Delta A, B + \Delta B)) \mid &\\ & \leq \eta\kappa_A ||x-y|| + \eta \kappa_B (||x-y|| + ||\phi -\psi||).
\end{align*}
\hfill$\square$
\end{remark}

As a consequence of Theorem~\ref{theorem:one}, we can recast~\eqref{eq:verifier-optimisation} as

\begin{equation}
    \label{eq:verifier-optimisationX}
        \begin{aligned}
\lambda^* = & \underset{x \in \mathscr{D}, \phi \in \Phi}{\min} && \lambda_{\textrm{min}}(\Xi_k(A, B)),
 \\
&~~~\mathrm{ s.t.}  && \eqref{eq:def-omega},  
        \end{aligned}
\end{equation}
and, exploiting Lemma~\ref{lemma:liplamdamin}, solve~\eqref{eq:verifier-optimisationX} to global optimality using a global Lipschitz solver. Note that the peculiar nature of $\Phi$ simplifies the resolution of \eqref{eq:verifier-optimisationX}, allowing us to split the verification problem in $p$ sub-problems with one loss of efficiency on $\phi$ at a time. Such problem formulation helps tackling the nonconvex nature of $\Phi$, which further complicates the verifier's task, and supports the solution via Lipschitz solvers, which are known to suffer from the curse of dimensionality.
Likewise, one can perform the verification process of~\eqref{eq:exp-Bemporad-schur} concurrently for all $j$ owing to the Lipschitz-continuity of the operator $\mathrm{min(\cdot)}$. It is therefore possible to evaluate the minimum eigenvalue of different $\Xi$ in parallel and select the smallest one as the value of the objective function for a given $x, \phi$.

\subsection{CEGIS-based procedure and analysis}\label{subsec:cegis}

\begin{algorithm}[!t]
	\caption{CEGIS-based PFTC learning method}\label{algo:OverallCegis} 
	\textbf{Initialization:} Set $\eta \ge \epsilon>0$, $\mc S_1 = (\hat A_1, \hat B_1)\in\Omega$ \\
	\smallskip
	\textbf{Iteration $(k \in \N_+)$:} 
	\begin{itemize}
		\setlength{\itemindent}{.2cm}
		\item[\small\circled{L}] Identify $\textrm{vert}(\textrm{conv}(\mc S_k))$
	\end{itemize}
        \smallskip
	\begin{itemize}
		\setlength{\itemindent}{.2cm}
		\item[\small\circled{L}] Solve \eqref{eq:learner-optimisation-LMI}, set $(P_k,K_k, H_k)$ as in \eqref{eq:solution-of-LMI}
		
		\textbf{If} \eqref{eq:learner-optimisation-LMI} \texttt{infeasible} \textbf{then} \texttt{exit}
	\end{itemize}
        \smallskip
	\begin{itemize}
		\setlength{\itemindent}{.2cm}
		\item[\small\circled{V}] Solve \eqref{eq:verifier-optimisationX}, $\forall j$, using Lipschitz global optimization	\\	
		\textbf{If} $\lambda^\ast \leq 0$ \textbf{:} $\mc S_{k+1}\leftarrow \mc S_k \cup \{(A^\ast,B^\ast)\}$, \texttt{repeat}\\
		\textbf{If} $\lambda^\ast > 0$ \textbf{:} $V=x^\top P_k x$, $K=K_k$, $H=H_k$, \texttt{exit}
	\end{itemize}
\end{algorithm}

\begin{figure}[t!]
	\centering
	\includegraphics[width=.98\columnwidth]{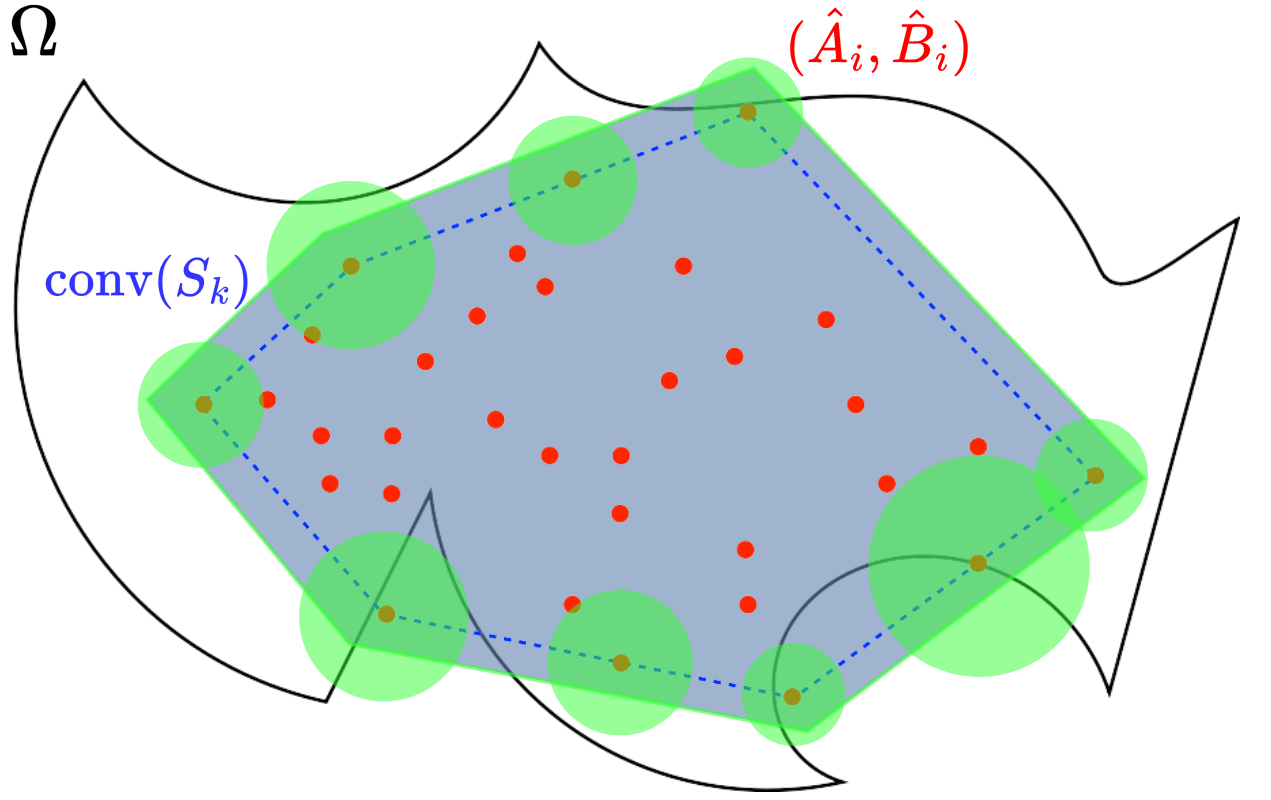}
	\caption{The region identified by the union of the green ``balls'' %
    centred on each vertex pair of matrices $\{(\hat A_i, \hat B_i)\}_{i=1}^{V_k}$ (red dots, among all the other samples) and the associated set of matrices with eigenvalues $(\epsilon/\lambda_{\textrm{max}}^2(P))$-distant from those belonging to $\textrm{cl}(\textrm{conv}(\mc S_k))$ (green line and its interior, with $\textrm{conv}(\mc S_k)$ represented by the blue dashed line), denotes the actual portion of volume where it is guaranteed that no counterexample can be found. As the iterations of Algorithm~\ref{algo:OverallCegis} progress, the resulting area will cover the whole of $\Omega$, regardless of its shape. }
	\label{fig:verifier-cegis-domain}
\end{figure}

Algorithm~\ref{algo:OverallCegis} reports the main steps of the proposed \gls{CEGIS}-based iterative scheme for the design of \gls{PFTC} policies based on Lyapunov arguments. 
Specifically, the {\tiny\circled{L}} bullets refer to tasks performed by the learner,
while the {\tiny\circled{V}} bullet to the one performed by the verifier. A intuitive depiction of the proposed procedure is shown in Fig.~\ref{fig:verifier-cegis-domain}, illustrating the geometrical intuition behind how a generic-shaped $\Omega$ is gradually covered by the polytope $\textrm{conv}(\mc S_k)$ and the ``safe balls'' spawning from every point in its frontier. 

Algorithm~\ref{algo:OverallCegis} may terminate with two possible outcomes, denoted by the \texttt{exit} commands.
First, if the verifier finds no counterexample, Algorithm~\ref{algo:OverallCegis} has converged and a control function $K$ is returned.
Alternatively, if the verifier returns a counterexample, the latter is added to the set $\mathcal{S}_{k+1}$ and the next learner iteration is started by solving a new instance of~\eqref{eq:learner-optimisation-LMI}. Our procedure is susceptible to the quality of the counterexample produced by the verifier, in turn associated to the choice of the hyperparameters of our method. 
Indeed, it may occur that an instance of~\eqref{eq:learner-optimisation-LMI} may fail to generate a controller (declaring infeasibility when solving~\eqref{eq:learner-optimisation-LMI}) due to the specific choice of said hyperparameters. In such a case, the user may want to restart the algorithm with a new choice of hyperparameters.
In case~\eqref{eq:learner-optimisation-LMI} remains feasible, Algorithm~\ref{algo:OverallCegis} enjoys finite-time convergence. This means that either the procedure declares infeasibility, or it returns a \gls{PFTC} policy within a finite-number of steps, as established hereby. 
\begin{theorem}
    Let $\eta \geq \varepsilon > 0$ and $(\hat{A}, \hat{B}) \in \Omega$. Then, in a finite number of steps Algorithm~\ref{algo:OverallCegis} either declares infeasibility or return a triplet $(Q, Y, Z)$ so that $V(x) = x^\top Q^{-1} x$ is a control Lyapunov function for \eqref{eq:model-linear-uncertain} with saturated input $u = \sat(Kx)$. 
\end{theorem}
\begin{pf}
    The proof is sketched by following the steps of \cite[Th.~2]{masti2023counter}. 
    From \eqref{eq:exp-Bemporad-schur} it follows that, at step $M$ of Algorithm~\ref{algo:OverallCegis}, the set $S^M \subseteq \Omega$, $S^M \eqdef \mathrm{conv}(\{(A_i,B_i)\}^M_{i=1})$ represents the set of matrices stabilised by $K_M$. We now prove that each new counterexample expands the set $S^{M}$ by (at least) a constant amount: as $\Omega$ is compact, and $S^M$ grows consistently at each iteration, the algorithm must terminate in a finite number of steps.  
 From Theorem~\ref{theorem:one} it follows that for any new counterexample $z_{M+1}=(\bar A_{M+1}, \bar B_{M+1})$ we have:
    \begin{align*}{}
    ||\bar A_{M+1}-A_s|| + & ||\bar B_{M+1}-B_s||
     > \frac{\varepsilon}{\eta},~\forall (A_s,B_s)\in S^M.
    \end{align*}
 Once assumed that~\eqref{eq:learner-optimisation-LMI} is feasible for all $M\ge0$ (otherwise Algorithm~\ref{algo:OverallCegis} terminates in a finite number of steps declaring infeasibility)   
 it holds that 
 $S^M \subset S^{M+1}\subseteq \Omega$. In turn, for a sufficiently large $k$, it will hold that $S^{M+k} \supseteq \Omega$, ensuring that the algorithm terminates in at most $M+k$ steps. 
 \qedsymbol
\end{pf}
 \begin{remark}
 From Remark~\ref{rem:JacLipSafeSet} and~\eqref{eq:LipSchitzGradient}, it also follows that for any tuple $[\bar x_{M+1}, \bar \phi_{M+1}]$ such that $z_{M+1}=(A|_{\bar x_{M+1}}=A|_{ x_{M+1}},B_{M+1}=B|_{\bar x_{M+1}, \bar \phi_{M+1}})$, 
 it will hold that
    \begin{align*}{}
    \kappa_A ||\bar x_{M+1}-y|| + &  \kappa_B (||\bar x_{M+1}-y|| \\
    &+ ||\bar \phi_{M+1} -\psi||) > \frac{\varepsilon}{\eta } 
    ~~~\forall [y,\psi] \in D^M,
    \end{align*}
    where $D^M \eqdef\{ [y,\psi] ~|~ y \in \mathscr{D},\psi \in \Phi , 
 (A|_y,B|_{y,\psi}) \in S^M\}$.
 Moreover, due to $f,g$ being $C^2$ and $S^M$ being a closed and connected set, we know that $D^M$ is closed and connected.
 From these considerations it follows that $D^{M+1} \supset D^{M}$ and therefore there exists a 
 sufficiently large $k$ such that $D^{k} \supseteq \mathscr{D} \times \Phi$. 
 \textit{}
 \hfill$\square$
 \end{remark}

Having now completed the theoretical derivations, we next discuss how the described method can be employed in realistic control case scenarios involving nonlinear dynamical systems subjected to a range of faults at actuators.

\section{Case study: PFTC for a hover-capable AUV}
\label{sec:experiments}

In this section, 
we test our \gls{IS-sat} controller on two realistic benchmarks, and compare our bespoke method against an $\mathcal{H}_\infty$ control, representing a common passive fault-tolerant alternative. 

The code relative to the \gls{IS-sat} control is written in Python 3.11, employing \texttt{cvxpy} as the optimisation library solving \eqref{eq:learner-optimisation-LMI}, while the verification task uses the Scipy SHGo library to solve the problem as formulated in \eqref{eq:verifier-optimisation}.
The experiments are run on a laptop computer with four 1.90GHz cores and 16 GB of RAM.

\subsection{A hover-capable AUV}
\label{sec:CEP_comparison}
Let us consider a hover-capable \gls{AUV}, which represents an \gls{AUV} operating in condition of neutral buoyancy, capable of maintaining prescribed position and attitude in presence of environmental disturbances. Hover-capable \glspl{AUV} are conventionally employed in surveillance of underwater maritime structures and in sonar-searches and mapping, where they are tasked with following a path at constant depth and constant forward speed. In order to follow a predefined path, the innermost control loop is designed to maintain the forward (surge) speed at a constant set value, to regulate the angular speed to zero, and tracking desired time-varying heading setpoints \cite{fenucci2024multi}. In such precise guidance and control applications, where \glspl{AUV} operate nearby the seabed carrying sensible and expensive instruments, faults at thruster can result in collisions with ground, or, in the worst case scenarios, in damages to the structures under surveillance or in the loss of the vehicle.  

\begin{figure}[t]
    \centering
	\includegraphics[width=0.8\linewidth]{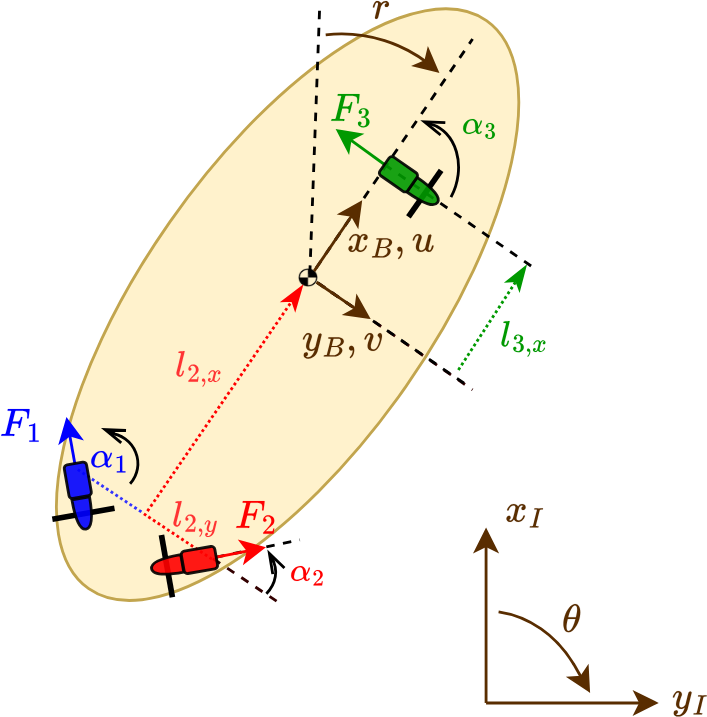}
	\caption{Hover-capable AUV with three (fixed) thrusters moving in the horizontal plane.}
	\label{fig:AUV-3t}
\end{figure}
      
To this aim, let us consider the two-state \gls{AUV} model from~\cite[Case study A]{grande2024passive}, representing a simplified version of an Autosub Hover \gls{AUV}. The vehicle mounts three fixed (i.e. non-rotating) thrusters, generating force along the thruster axis in both the positive and negative directions, as illustrated in Fig.~\ref{fig:AUV-3t}. Such vehicle is designed to operate near infrastructures at low speeds and constant depth, while retaining the capacity to control pitch and yaw angle \cite{fenucci2024multi}. 
The equations of motions describing the vehicle dynamics are derived from the robot-like vectorial model for marine crafts \cite{fossen2011handbook}, 
while restricting our analysis to the surge and yaw rate degrees of freedom, resulting in\footnote{For a detailed derivation of the model, we refer the interested reader to~\cite[Ch. 4, 6.5.2]{grande2024actuator}}: 
\begin{align}
	\label{eq:CEP-AUBV}
	\left\{\begin{array}{ll}
\dot{x}_1=& \dfrac{-X_ux_1-X_{uu}x_1^2+\phi_1 F_{1,x}+\phi_2F_{2,x}+\phi_3F_{3,x}}{m}	\\	\dot{x}_2=& \dfrac{-N_r x_2 - N_{rr}x^2_2+ (-F_{1,x}l_{1,y}+F_{1,y}l_{1,x})\phi_1}{J_z}\\
&+ \dfrac{(-F_{2,x}l_{2,y}+F_{2,y}l_{2,x})\phi_2}{J_z}\\
&+ \dfrac{(-F_{3,x}l_{3,y}+F_{3,y}l_{3,x})\phi_3}{J_z}
	\end{array}\right.,
\end{align}
where $x \eqdef [x_1, x_2]^\top$ denotes the surge speed and angular acceleration about the vehicle vertical axis, and $u \eqdef [F_1, F_2, F_3]^\top$ denotes the control input encompassing the forces generate by the three thrusters, namely the aft port ($F_1)$, aft starboard ($F_2$) and bow thruster ($F_3)$. Moreover, the model includes $u \in [-\bar u,\bar u]$, where $\bar u$ denotes the saturation value of the thrusters and $F_{i,x}=F_i \sin(\alpha_i)$ and $F_{i,y}=F_i \cos(\alpha_i)$ represent the projections of the thruster force $F_i$ along the $x_b$ and $y_b$ body-axes, with the convention reported in Fig.~\ref{fig:AUV-3t}. Finally, $X_u$, $X_{uu}$ denote the linear and quadratic surge drag coefficients, while $N_r$, $N_{rr}$ the linear and quadratic yaw drag coefficients and $\phi_i$ the efficiency associated to the $i$-th thruster.

Next, we detail the design of the control laws. The state domain is defined as $\mc D=[-2,2]^{2}$, 
which extensively covers the usual range of linear and angular velocities of the hoover-capable \glspl{AUV}. In this case study, the control saturation value is set to $\bar u=38.0$ N for all actuators, value selected within the conventional saturation range of common underwater thrusters, such as the BlueRobotics T200 thrusters. 
For our synthesis method, we consider a discrete-time version of model~\eqref{eq:CEP-AUBV}, which is discretized via the explicit Euler method with a time step of $0.01$ s. %
By setting 
$\eta=50$, $\varepsilon=10^{-4}$, $\tau = 0.999$, we are able to synthesise a sound controller within $7$ iterations of Algorithm~\ref{algo:OverallCegis} (with an overall run time of a few seconds), resulting in the following feedback gain
\begin{equation}
    K_\mathrm{IS-sat} = 10^3 \begin{bmatrix}
		-43.987 &    0.308
		\\
		-30.985 &   7.948
		\\
		-1.187 &   37.481
	\end{bmatrix},
\end{equation}
which is in turn applied in closed-loop as 
\begin{align}\label{eq:sat_con_scenario1}
	u(t)=38\cdot\sat_{\mathscr U}(K_\mathrm{IS-sat}  e(t)).
\end{align}

Recall that the \gls{IS-sat} method represents an optimised version of a classical polytopic formulation for uncertain systems. 
For comparison, the standard polytopic formulation from~\cite{rubin2020interpolation}, under the same modeling uncertainties, yields an optimization problem with $2^{13}$ constraints due to the 10 time-varying coefficients of the $A$, $B$ matrices and the 3 control signals, which is beyond the computational requirements of a common office laptop\footnote{We provide further computational details and comparisons in the more extensive case study reported in Section~\ref{sec:AUV_5dimensional}}.

\textbf{Performance comparison: }
The proposed approach is tested against two $\mathcal{H}_\infty$ control laws, which represent a conventional framework to solve PFTC problems in \gls{AUV} applications (see, for instance, ~\cite{kaminer1991control, grande2024passive}). 
As in~\cite{grande2024passive}, we formulate the $\mathcal{H}_\infty$ optimisation problem as the simultaneous stabilisation of four operational modes, namely the faultless mode and the three modes characterised by one (distinct) thruster failing. Each mode is obtained by linearising equation~\eqref{eq:CEP-AUBV} around $\bar x=[0.5,0]$ and by selecting $\phi_1 = \phi_2 = \phi_3 =1$ in the faultless mode, and $\phi_i=0$ for the mode corresponding to the $i$-th thruster at fault. %
To devise controllers with different performance characteristics, two optimisation problems are formulated. A target response time of 10 s (a realistic value for the application being considered) is set as a design constraint (hard goal) for both the optimisation problems, while the maximum tolerated steady-state error is set as objective (soft goal) and varied between the two controllers. The first control law, defined `aggressive' and denoted as $\mathcal{H}_\infty^a$, is designed to ensure a low maximum steady-state error---set at 5\%---assuming that energy and power consumption are not the key design driver. The second `conservative' control law, denoted as $\mathcal{H}_\infty^c$, is designed relaxing the maximum steady-state error at 40\% and by imposing a saturation on the actuator as soft optimisation constraint. The synthesised controllers result in
\begin{equation}\label{eq:Hinf_agg}
    \mathcal{H}_\infty^a = 10^4 \cdot \begin{bmatrix}
		-0.2400 &    -0.2865
		\\
		-0.2553 &   0.3037
		\\
		0.0013 &   2.2510
	\end{bmatrix},
\end{equation}

\begin{equation}\label{eq:Hinf_cons}
    \mathcal{H}_\infty^c = \begin{bmatrix}
		-232.1081 &  -277.2772
		\\
		-183.4074 & 219.4158
		\\
		0.0298 & 776.4082
	\end{bmatrix}.
\end{equation}

\begin{figure}[t]
    \includegraphics[width=8.4cm]{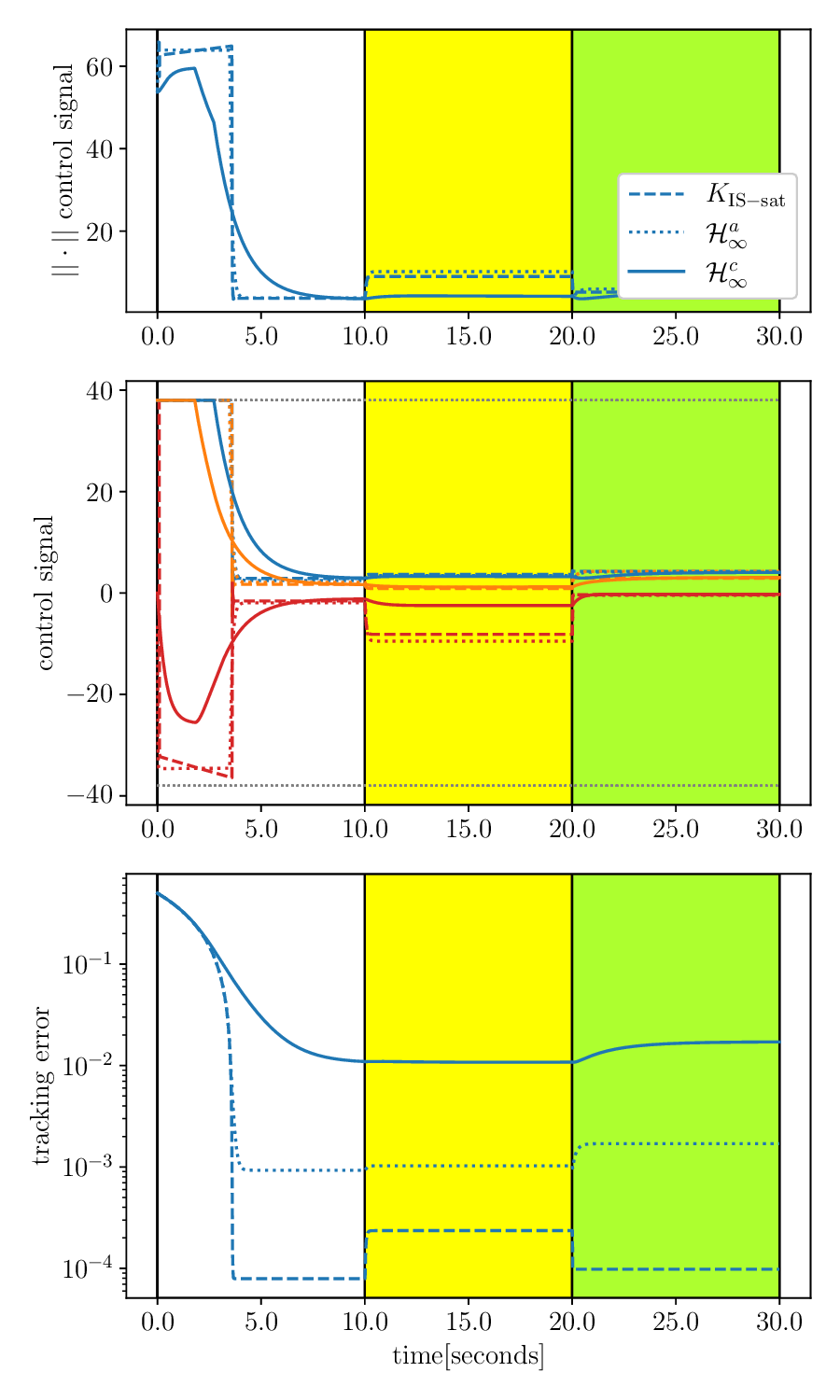}
    \caption{Comparison of our proposed controller $K_\mathrm{IS-sat}$ with the two $\mathcal{H}_\infty$ controllers for tracking a constant reference $\bar{x} = [0.5, 0]$. The top plot shows the norm of the control input over time. The middle plot shows the control action applied to each input channel, with gray lines indicating saturation limits. The bottom plot shows the norm of the tracking error (on both components). The three vertical coloured regions indicate the three fault modes. %
    }
    \label{fig:AUV-CEP}
\end{figure}

Next, we evaluate the performance of the three control laws by tracking the reference point $\bar{x} = [0.5, 0]$ over a simulation horizon of $T = 30$ s. The simulated scenario is divided into three phases to represent different operational modes. In the first section of the simulation ($0 \leq \text{t} \leq 10$ s), the system operates without faults. In the second phase ($10 < \text{t} \leq 20$ s), thruster $F_3$ operates at 10\% of its nominal efficiency, while the other thrusters operate nominally. Finally, in the last third of the scenario ($20 < \text{t} \leq 30$ s), $F_2$ operates at 10\% of its nominal efficiency, while the other thrusters operate nominally.
Figure~\ref{fig:AUV-CEP} shows the simulation results in terms of the magnitude of the control effort and tracking error (it is recalled that even if the $\mathcal{H}_\infty$ controllers do not explicitly exhibit the saturation limits within the ~\eqref{eq:Hinf_agg} and \eqref{eq:Hinf_cons}, the physical limits of the actuators are accounted for within $u$ in model~\eqref{eq:CEP-AUBV}). It can be noted that our controller exhibits similar closed-loop performance to the `aggressive' $\mathcal{H}_\infty^a$ control design in terms of tracking error. %

\begin{figure}[t]
    \includegraphics[width=\linewidth]{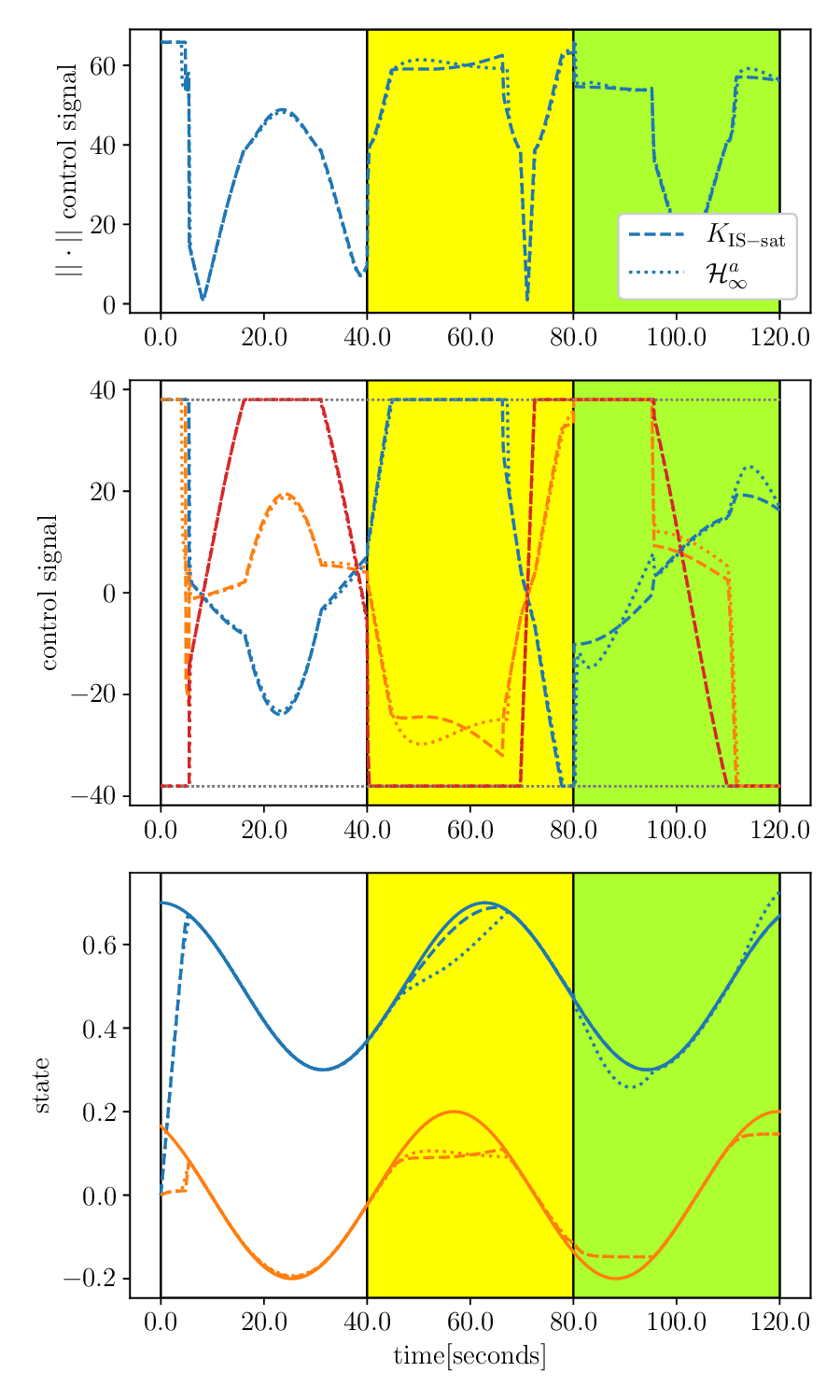}
    \caption{Comparison of the our controller $K_\mathrm{IS-sat}$ with the aggressive $\mathcal{H}^a_\infty$ controller from~\cite{grande2024passive} for tracking a sinusoidal reference signal. In the third plot, the solid line represents the reference trajectory, with $x_1$ in blue and $x_2$ in orange. %
    }
    \label{fig:AUV-sine}
\end{figure}

Next, we further investigate the performance of our controller and of the $\mathcal{H}_\infty^a$ controller while tracking sinusoidal references on both the $x_1$ and $x_2$. We report the result in Fig.~\ref{fig:AUV-sine}, illustrating a simulation with a time horizon $T = 120$ s. In this case, \gls{IS-sat} also demonstrates tracking performance at least as good as $\mathcal{H}_\infty^a$.

\begin{figure}[t]
    \centering
    \includegraphics[width=8.4cm]{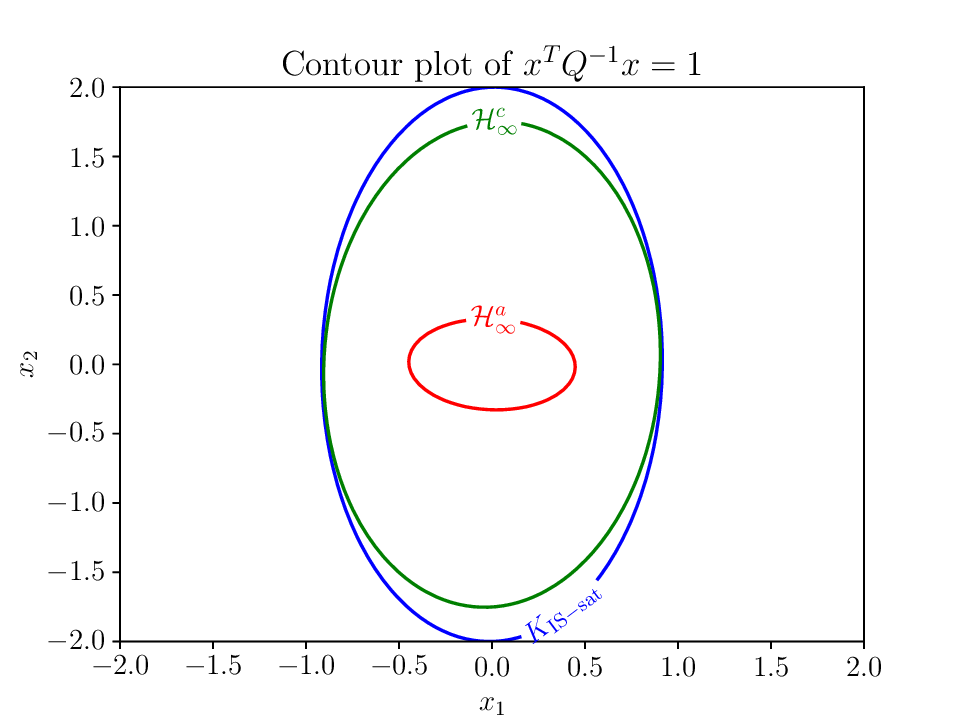}
    \caption{Domains of attraction (computed according to~\cite{rubin2020interpolation}), associated to the $\mathrm{IS-sat}$ and the two $\mathcal{H}_\infty$ controllers from~\cite{grande2024passive}. The proposed controller achieves the largest domain of attraction. This is particularly noteworthy given its comparable performance to the aggressive $\mathcal{H}_\infty^a$ controller, which exhibits significantly smaller domains of attraction.}
    \label{fig:AUV-Basin}
\end{figure}

Remarkably, our controller enjoys significantly better theoretical robustness properties, which can be observed by comparing the domains of attraction for the three feedback laws. To compute these, we solve Problem~\eqref{eq:learner-optimisation-LMI} with the additional constraint $KQ = Y$, still using the CEGIS loop as suggested in~\cite{rubin2020interpolation}.
The resulting domains of attraction are shown in Fig.~\ref{fig:AUV-Basin}. As expected, our approach achieves the largest domain of attraction among the three control laws. %
Notably, our approach also compares well with the neural network-based approach proposed in~\cite{grande2024passive}, which has a domain of attraction equal to a sphere of radius 1 (not depicted in Fig.~\ref{fig:AUV-Basin}).
It is important to note that these domains of attraction are very conservative estimates of the actual domains. Nevertheless, especially when considered in conjunction with the results in Fig.~\ref{fig:AUV-CEP}, they demonstrate that our technique is not significantly constrained by the technical constraints in Equation~\eqref{eq:learner-optimisation-LMI}, which are introduced to enhance the convergence of the control synthesis.

\subsection{A 5-dimensional AUV model}\label{sec:AUV_5dimensional}

\begin{figure*}[h!t]
	\normalsize
	\begin{align}
		\left\{\begin{array}{ll}
			\dot{x}_1=& \dfrac{-X_ux_1-X_{uu}x_1^2+mx_2x_3+\phi_1F_{1,x}+\phi_2F_{2,x}+\phi_3 F_{3,x}+\phi_4F_{4,x}}{m}	\\
			\dot{x}_2=& \dfrac{-Y_vx_1-Y_{vv}x_1^2-mx_1x_3+\phi_1F_{1,y}+\phi_2F_{2,y}+\phi_3F_{3,y}+\phi_4F_{4,y}}{m}	\\			
			\dot{x}_3=& \dfrac{-N_r x_2 - N_{rr}x^2_3+ (-F_{1,x}l_{1,y}+F_{1,y}l_{1,x})\phi_1}{J_z}\\
			&+ \dfrac{(-F_{2,x}l_{2,y}+F_{2,y}l_{2,x})\phi_2+(-F_{3,x}l_{3,y}+F_{3,y}l_{3,x})\phi_3+(-F_{4,x}l_{4,y}+F_{4,y}l_{4,x})\phi_4}{J_z}\\
			\dot{x}_4 =&x_3\\
			\dot{x}_5 =&x_4
		\end{array}\right.
		\label{eq:5thOrderSys}
	\end{align}
	\hrule
\end{figure*}

Let us now analyse the performance of a four-dimensional, higher-fidelity version of the \gls{AUV} model employed in \S \ref{sec:CEP_comparison}. Besides the previous two state variables, the model now accounts for the angular rate about the vertical axis ($x_3$) and for the yaw angle ($x_4$). %
Introducing the angular rate dynamics allows to account for the Coriolis forces, which introduce cross-coupling terms between the degrees of freedom, in turn increasing the nonlinearity of the model. Finally, a fifth state (without direct physical meaning) is introduced as the integral action of the yaw angle, to ensure zero-error tracking for constant headings setpoints~\cite[\S IV-A]{zanon2021constrained}. 

\begin{figure}[ht]
	\centering
	\includegraphics[width=.9\columnwidth]{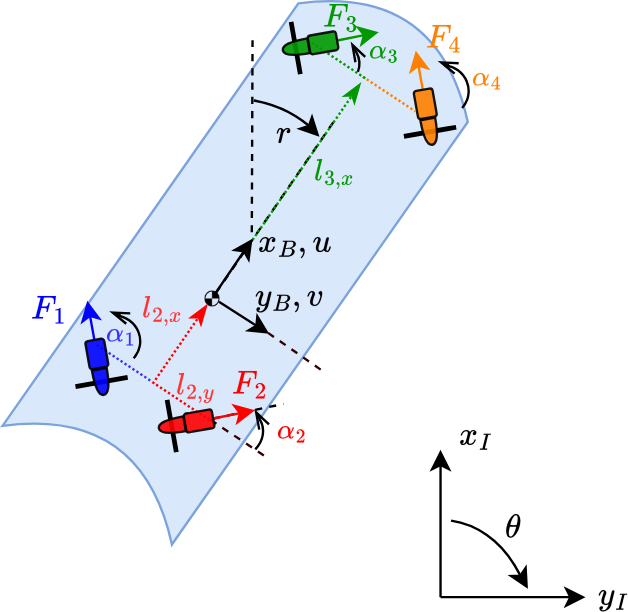}
	\caption{Hover-capable AUV with four (fixed) thrusters moving in the horizontal plane.}\label{fig:auv_model}
\end{figure}

For this study, we employ an \gls{AUV} with four thrusters in a X-shape configuration \cite{baldini2018model}. Each of the four non-azimuthing (i.e. not rotating) thrusters, two located aft and two stern of the centre of gravity, can generate positive and negative thrust force. Such a vehicle is illustrated in Fig.~\ref{fig:auv_model}, with the overall model dynamics reported in~\eqref{eq:5thOrderSys} (derived from ~\cite[Ch. 7.3]{grande2024actuator})%
. 

It should be noted that the nonlinear part of the \gls{AUV} dynamics~\eqref{eq:5thOrderSys} includes $p=3$ input variables (namely $x_1, x_2,$ and $x_3$) leading to a total number of 21 time-varying uncertain elements ($A \in \mathbb{R}^{3 \times 3}$, $B \in \mathbb{R}^{3 \times 4}$). %
The approach enumerating all vertices of a hypercube-shaped $\Omega$ accounting for 4 possible faults encompasses $2^{21} \times 2^4$ constraints, which is clearly beyond the capabilities of most of the modern hardware.

To synthesise the $K_\mathrm{IS-sat}$ controller, we start by selecting a state domain as $\mc D=[-2,2]^{3}$ for the first three components of the state vector, whilst we do not restrict the remaining two variables (namely the integral variables): this domain choice well exceeds the usual range of dynamics encountered by slow moving vehicles, which dynamics are usually bounded under 1 m/s in the linear speeds and under 0.18 rad/s (i.e. $\approx$10 deg/s) in yaw rate. The saturation threshold is once again set to $\bar u=38.0$ N for all actuators to resemble realistic commercial thruster values such as the T200. 
For this case study, we employ the same hyper-parameters values of the previous test, namely $\eta=50$, $\varepsilon=10^{-4}$, $\tau = 0.999$, and achieves convergence in 8 CEGIS loop iterations.

\begin{figure*}[h!t]
	\normalsize
    \begin{align}\label{eq:sat_con_2}
    	u(t)=38.0\ \sat_{\mathscr U}(\begin{bmatrix}
    		-50.93 &    -48.31 & 592.27 & 114.93 & 5.49
    		\\
    		45.28 & -45.56 & 540.61 & 107.57 & 5.14
    		\\
    		46.77 & -47.90 & -540.79 & -104.30 & -4.98
            \\
            -41.89 & -44.06 & -492.74 & -96.09 &-4.59
            \end{bmatrix} e(t)).
    \end{align}
\end{figure*}

\begin{figure}[t]
	\includegraphics[width = 8.4cm]{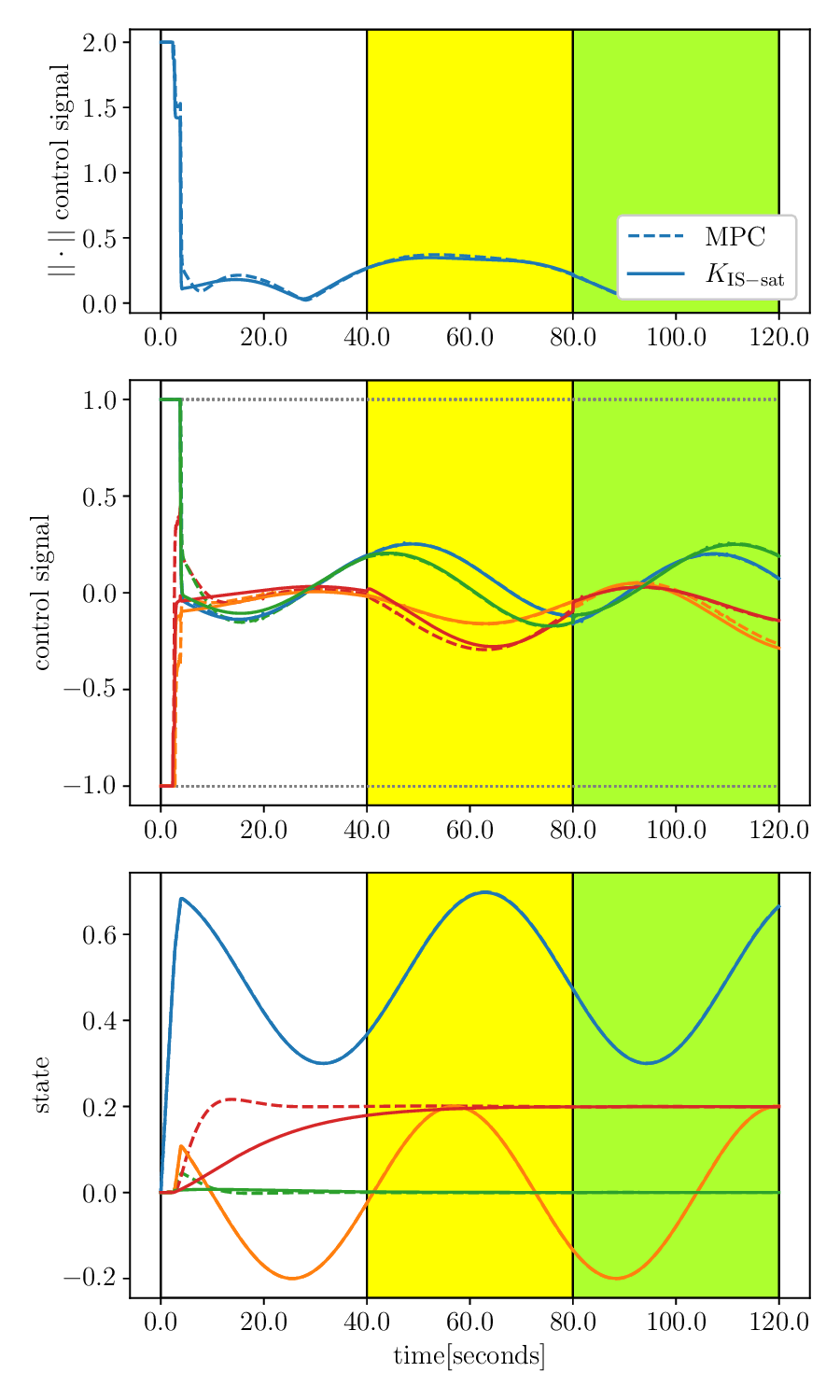}
	\caption{Comparison of \gls{IS-sat} with a nonlinear $\mathrm{MPC}$, with model dynamics initialised to $x=0$. Control signal plots have been normalised in the $[-1,1]$ range for ease of readability. State plot legend: $x_1$ in blue, $x_2$ in orange, $x_3$ in green, $x_4$ in red; solid line associated to $K_\mathrm{IS-sat}$, dashed line to MPC.    
    It can be appreciate how \gls{IS-sat} and MPC have comparative tracking performance for $x_1$ and $x_2$ (the solid and dashed lines superimpose), while MPC tracks $x_4$ 6x faster.}
	\label{fig:5states_Comparison}
\end{figure}

\begin{figure}[t]
	\includegraphics[width=8.4cm]{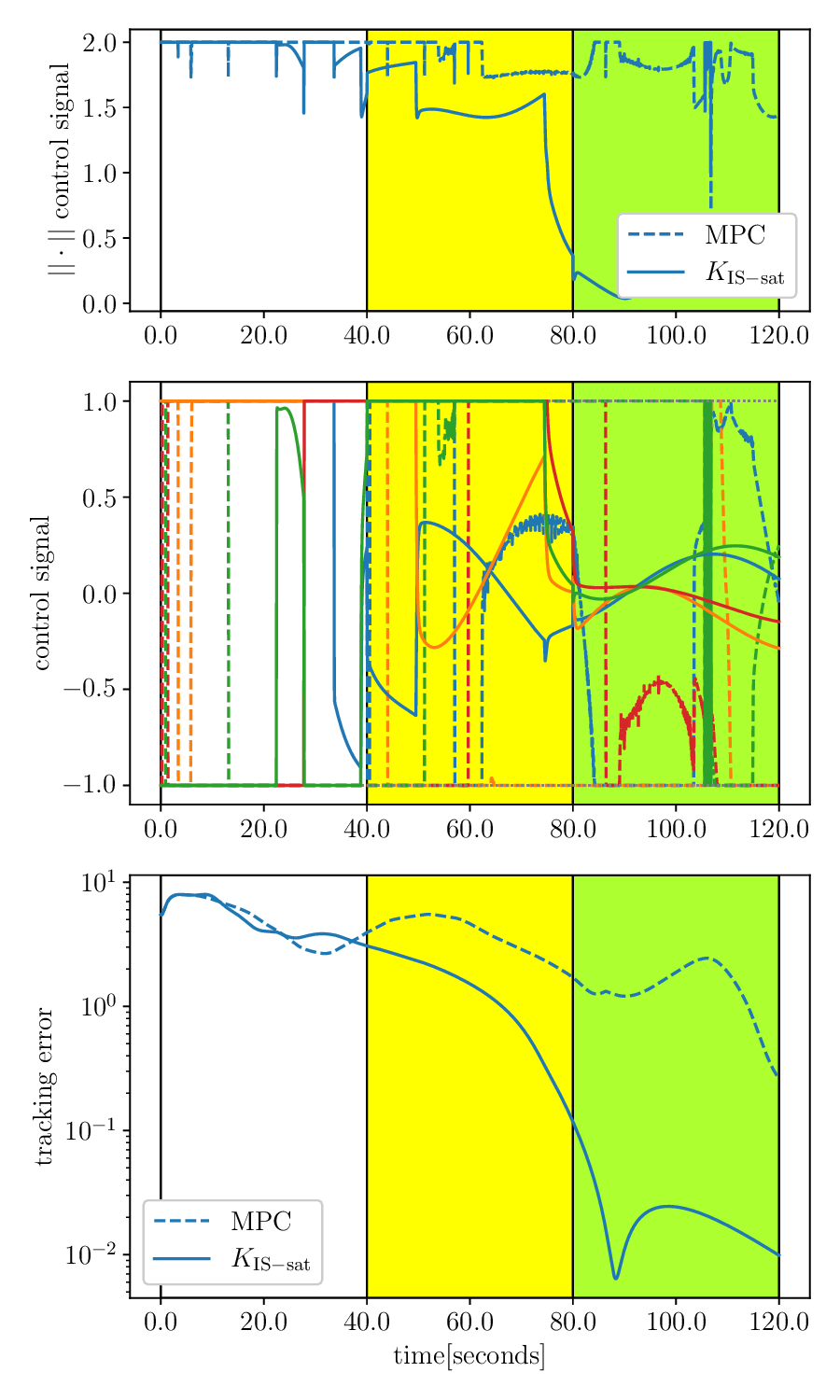}
	\caption{Comparison of \gls{IS-sat} with a nonlinear $\mathrm{MPC}$, with model dynamics initialised to $x=2$, and with the same reference signal of Fig.~\autoref{fig:5states_Comparison}. %
    Control signal plots have been normalised in the $[-1,1]$ range for ease of readability. State plot legend: $x_1$ in blue, $x_2$ in orange, $x_3$ in green, $x_4$ in red; solid line associated to $K_\mathrm{IS-sat}$, dashed line to MPC. It can be appreciate how, upon convergence, \gls{IS-sat} tracks the reference values much more accurately. %
    }
	\label{fig:5states_ComparisonFrom2}
\end{figure}

Let us now compare our control gain $K_\mathrm{IS-sat}$ against a fully fledged nonlinear MPC scheme built using~\eqref{eq:5thOrderSys} in no-fault condition, which represents the \emph{de facto} standard approach to control dynamical systems accounting for state and input constratints. %
We adopt a conventional LQR-like cost function~\cite[Ch. 8.4]{duan2013lmis} with $Q=1$ and $R=10^{-3}$, and we set the prediction horizon to 50 steps. %
As a first test, the \gls{AUV} is set to track two sinusoidal references on $x_1, x_2$, while tracking a constant yaw reference of $\bar x_4=0.2$. As in case study~\ref{sec:CEP_comparison}, we consider three sequential phases consisting of a no-fault phase, followed by a phase where the second actuator ($F_2$) functions at 10$\%$ efficiency to terminate with a phase where the third actuator ($F_3$) stands at 10$\%$ efficiency. Next, we consider two different initial conditions and perform two tests: $a)$ the simulation starts at the origin, illustrated in Fig.~\ref{fig:5states_Comparison}; 
$b)$ the simulation starts at the corner of our domain, namely $x(0)=[2,2,2,2,2]^T$, reported in Fig.~\ref{fig:5states_ComparisonFrom2}. 
As it can be appreciated in the state tracking plot within Fig.~\ref{fig:5states_Comparison}, both $K_\mathrm{IS-sat}$ and MPC are capable of satisfactory performance in tracking the sine signals on $x_1, x_2$, whilst the regulation of $x_4$ is much faster for the MPC controller with respect to our proposed approach (10 s vs 60 s).
However, when the initial condition is at the boundary of the domain, the MPC approach behaves rather poorly compared to \gls{IS-sat}, as illustrated in the tracking error graph reported within Fig.~\ref{fig:5states_ComparisonFrom2}. %
Moreover, the computational burden of the MPC approach is much more demanding compared to our method's: simulations involving MPC required $\approx 40$ s vs a negligible runtime required by the \gls{IS-sat} static controller (i.e. $< 1$ s).
It follows that our controller is promptly available to be deployed in embedded applications running on unsophisticated hardware with limited computational budget. In the specific case under analysis, IS-sat required $< 1$ MB of RAM, while the MPC\footnote{The MPC is implemented in JAX with Just-In-Time compilation and with a prediction horizon of 50 steps.} requires $\approx 285$ MB of onboard RAM, with a consequent increased energy demand, which is at premium in real underwater embedded applications. 

\subsection{Validation of the control in the OpenMAUVe simulator}

In this section, we focus on further enhancing the trustworthiness of the proposed control method. More specifically, we test and discuss the performance of the control law designed in \S \ref{sec:AUV_5dimensional} in a higher fidelity simulation environment. 

Hover-capable \glspl{AUV} can be tasked with autonomous inspection of monopiles, representing the submerged structures of the offshore wind turbines laying into the seabed \cite{pedersen2024towards}. Routine inspection of monopiles is of paramount importance to monitor the structure degradation due to corrosion and biofouling, which can lead to catastrophic outcomes if advancing undetected \cite{sindi2024advancing}. \glspl{AUV} represent the ideal candidates to effectively scale operations as a growing number of wind farms are being installed. A monopile inspection is conventionally performed by an \gls{AUV} carrying a front-looking camera, moving sideways around the monopile, keeping the camera (i.e. the vehicle fore end) pointed towards the monopile. 

This case study is performed within the OpenMAUVe simulator environment \cite{grande2025openmauve}. OpenMAUVe allows to simulate the underwater dynamics of \glspl{AUV}, accounting for both nonlinear cross-coupling and nonlinear effects associated to the hydrostatics and the hydrodynamics of fully submerged vehicles in motion. For the present study, a hover-capable neutrally buoyant vehicle is designed within the simulator, with viscous effects encompassing skin friction, form damping and added mass effects kept into account. Four thrusters are then added to the vehicle in accordance with the actuator scheme previously depicted in Fig.~\ref{fig:auv_model}. 

Next, we focus on designing the overall onboard autonomous architecture as composed of two elements: $a)$ an overarching \emph{guidance} law; $b)$ a \emph{low-level} control law. The guidance law is in charge of defining a set of waypoints that, starting from an initial random location, allows the \gls{AUV} to approach the monopile and to perform the inspection routine by revolving around the structure. The guidance law produces the set of reference speeds and heading, which is passed to the low-level control. In turn, the low-level control, which we focus on in this article, generates the forces to track the desired reference values. For this case study, we select a line of sight-based guidance law which allows the \gls{AUV} to follow an hexagonal path around the monopile, resembling a realistic mode of operations (different guidance laws could be selected \cite{breivik2008guidance}: for the scope of this work, we are interested in any guidance law which provides realistic time-varying references).

\begin{figure}[ht]
    \centering
    \includegraphics[width=\linewidth]{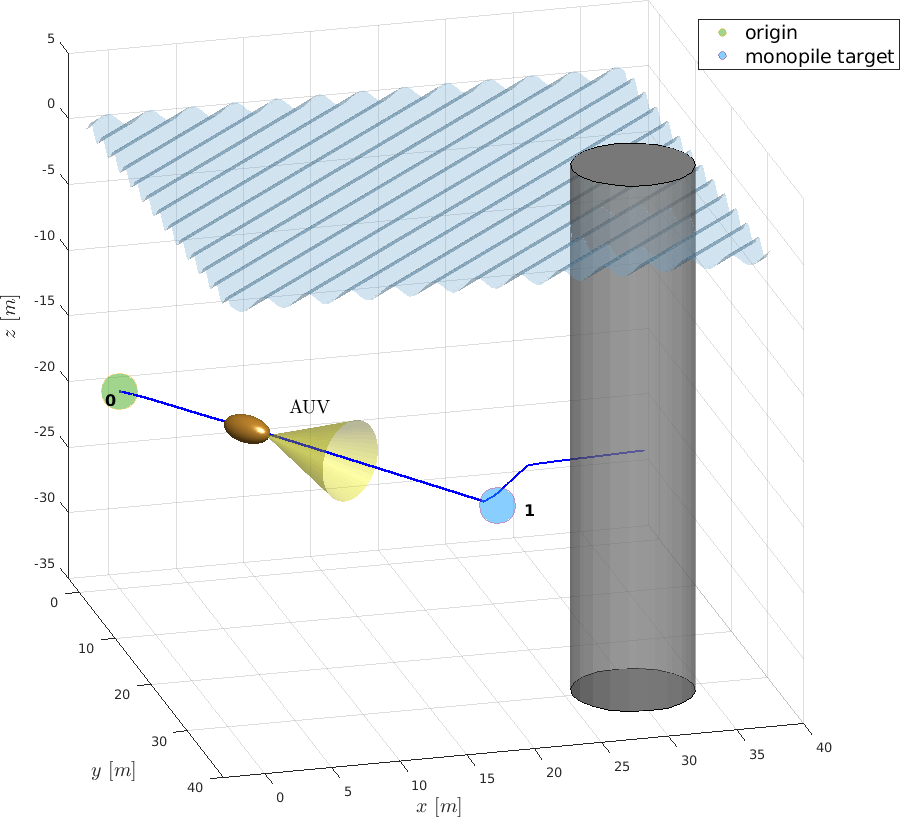}
	\caption{AUV (in orange) moving from an initial random position (waypoint 0) to a location close to the submerged monopile (waypoint 1). The AUV hovers at a depth of 20 m, mounting a forward looking camera, with the camera viewing cone displayed in yellow.}
	\label{fig:AUV-path_reach}
\end{figure}

\begin{figure}[ht]
    \centering
    \includegraphics[width=\linewidth]{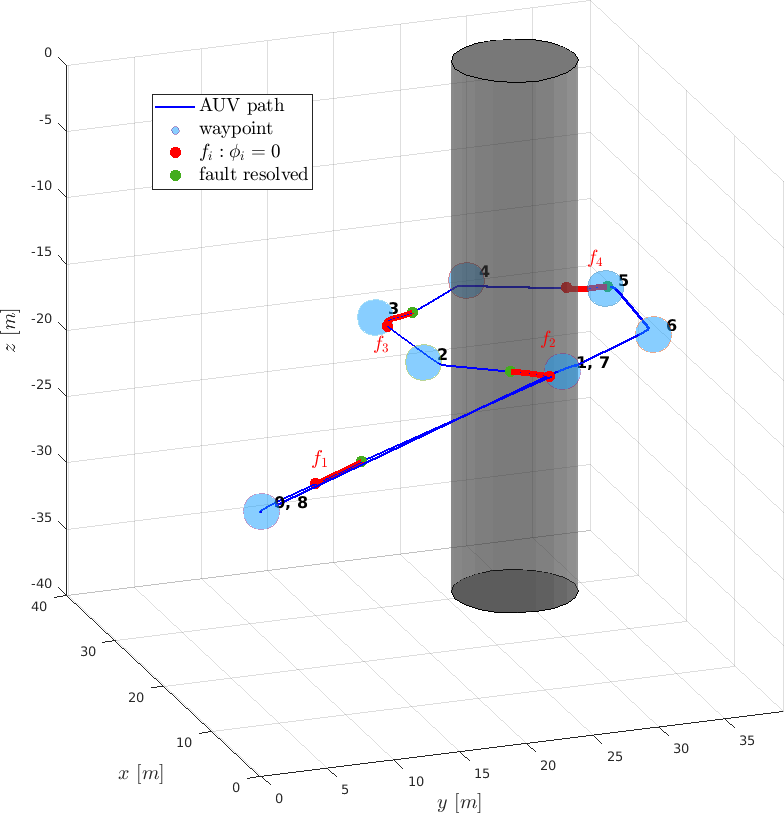}
	\caption{AUV overall path (in blue) with the segments characterised by faults at actuators (in red).}
	\label{fig:AUV-path_with_faults}
\end{figure}

The first segment of the \gls{AUV} path is defined such that the vehicle transitions from an initial random position (waypoint 0) to the closest point at a fixed distance of 5 m from the monopile (waypoint 1). Such an initial (approaching) phase is illustrated in Fig.~\ref{fig:AUV-path_reach}. 
The overall \gls{AUV} path is illustrated in Fig.~\ref{fig:AUV-path_with_faults}: once the \gls{AUV} terminates the inspection of the monopile (i.e. reaching waypoint 7), a commanded to return to the original location (waypoint 8) is sent. During the operation, the vehicle undergoes four fault modes, each one associated to the fault of one (unique) thruster. To verify the reliability of the proposed controller, we test the worst case scenarios, namely we test $f_i: \phi_i=0$ for $i \in [1, 4]$, the mode corresponding to the $i$-th thruster complete rupture (0\% of efficiency, while the other three thrusters work nominally). The faults are injected both during the reaching phase and during the inspection phase, with the faults lasting for 60 s each. The simulation automatically terminates after 2295 s, when the \gls{AUV} reaches the terminal waypoint; overall, the scenario takes about 2 s to be simulated.  

\begin{figure}[ht]
    \centering
    \includegraphics[width=\linewidth]{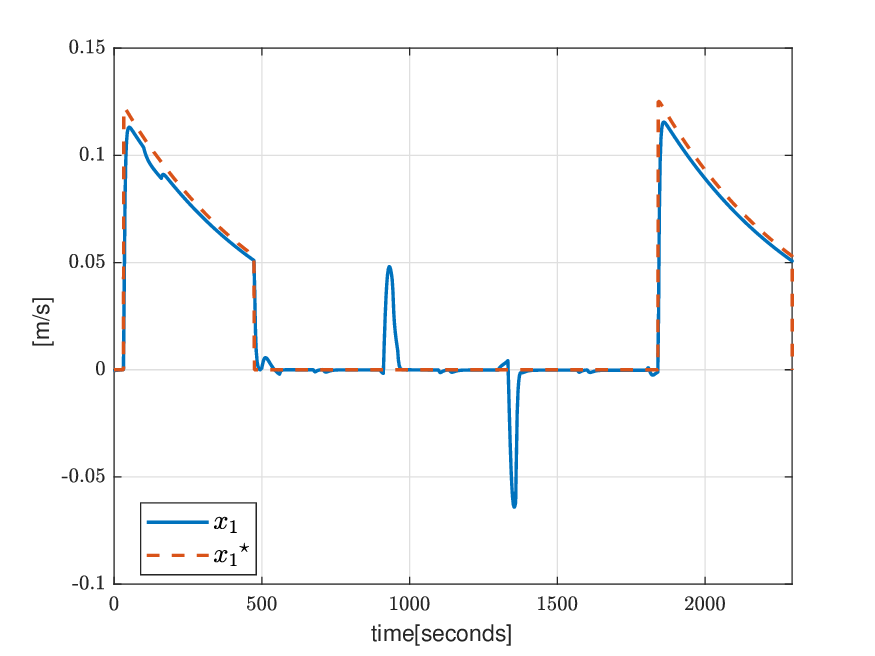}
	\caption{AUV surge speed ($x_1$) reference tracking with the control law \gls{IS-sat}~\eqref{eq:sat_con_2} during the path illustrated in Fig.~\ref{fig:AUV-path_with_faults}.}
	\label{fig:AUV-tracking_surge}
\end{figure}

\begin{figure}[ht]
    \centering
    \includegraphics[width=\linewidth]{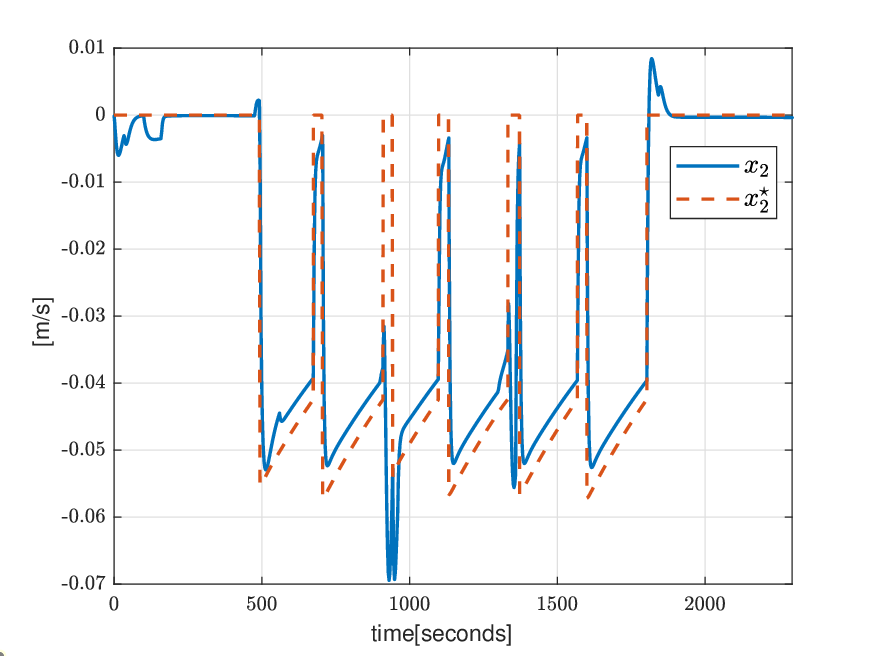}
	\caption{AUV sway speed ($x_2$) reference tracking with the control law \gls{IS-sat}~\eqref{eq:sat_con_2} during the path illustrated in Fig.~\ref{fig:AUV-path_with_faults}.}
	\label{fig:AUV-tracking_sway}
\end{figure}

\begin{figure}[ht]
    \centering
    \includegraphics[width=\linewidth]{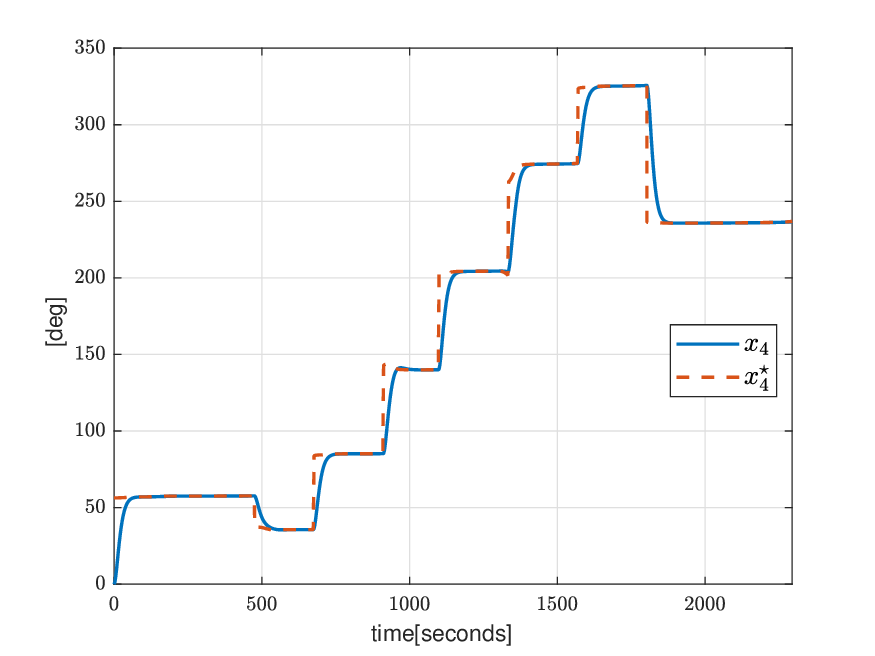}
	\caption{AUV yaw angle ($x_4$) reference tracking with the control law \gls{IS-sat}~\eqref{eq:sat_con_2} during the path illustrated in Fig.~\ref{fig:AUV-path_with_faults}.}
	\label{fig:AUV-tracking_yaw}
\end{figure}

Finally, we analyse the tracking performance of the control law~\eqref{eq:sat_con_2} in the aforementioned scenario. We specifically focus on $x_1$, $x_2$, $x_4$, as surge speed, sway speed and yaw angle are the driving variables steering the \gls{AUV} during the path tracking (the references for $x_3$ and for $x_5$ are set to 0) \cite{fossen2011handbook}. The reference surge speed is illustrated in Fig.~\ref{fig:AUV-tracking_surge}, the sway speed in Fig.~\ref{fig:AUV-tracking_sway} and the yaw angle tracking in Fig.~\ref{fig:AUV-tracking_yaw}.
As it can be seen, the guidance law provides a time-varying surge speed $x_1^\star = 0.12$ m/s during the reaching phase (with $x_2^\star = 0.0$ m/s), while it requires a reference $x_2^\star = -0.055$ m/s (with $x_1^\star = 0.0$ m/s) during the inspecting phase (the negative sign denoting that the AUV is requested to move port). Both reference values are gradually decreased as the \gls{AUV} approaches the next waypoint to slow down the vehicle, in turn allowing for a smoother transition to the next waypoint. As it can be noticed in Fig.~\ref{fig:AUV-path_with_faults}, the yaw angle reference is updated 8 times, corresponding to the requests to switch to the next waypoint. Despite the four different faults being injected, the vehicle is still able to accurately track to required setpoints. 
Although such occurrence is highly unlikely to occur in practical operations, we purposely time the faults to occur when the guidance law is switching waypoint. The worst case tracking performance is in fact recorded in one such occasion: as it can be seen in Fig.~\ref{fig:AUV-tracking_sway}, the worst case performance occur in the interval 920-945 s, when a relative error of 36\% is recorded on the sway speed. Such high error is due to the fault injected just before waypoint 3 is reached, namely when the control law was already attempting to compensate for the change in actuator dynamics. Even in this case scenario, it can however be seen that the control architecture is able to compensate for the fault, proceeding to reach the next target waypoint and eventually autonomously terminating the mission profile.

\section{Conclusion}
\label{sec:conclusion}

Capitalizing on Lyapunov arguments and LMI reformulations, we have presented an automated method to design \gls{PFTC} laws tailored to nonlinear systems affected by both partial and total actuators faults, including actuator saturation and time-varying tracking capability.
Our \gls{IS-sat} tackles nonlinear systems via translation to linear parameter-varying models, whose uncertainty set is bounded for models accounting for actuator faults.
By exploiting a counterexample-based approach, our technique is able to outperform conventional techniques both from the realm of robust control (i.e., $\mathcal{H}_\infty$) and from the nonlinear systems literature (i.e., nonlinear MPC). Notably, with respect to other counterexample-based techniques, our technique is guaranteed to converge within a finite number of iterations, either providing a control solution or declaring infeasibility. The proposed control approach provides
encouraging results for experimentation with higher dimensional systems, which are usually problematic with automatic synthesis techniques.
Moreover, since \gls{IS-sat} consists of static gain, it can be easily implemented in embedded applications where energy consumption is critical, or where limited computational resources are allocated to the control subsystem. In the presented study, the nonlinear MPC requested $\approx$285x the RAM usage of \gls{IS-sat}. Our approach is applied to the control of \glspl{AUV}, but is relevant to every control application where additional sensors or fault monitoring algorithms can not be designed, implemented, or run in real-time due to energy, hardware or cost constraints. Future work will focus on studying the scalability to more complex, higher-dimensional models.

\printglossary
\printglossaries

\bibliographystyle{plain}        %

\end{document}